\PassOptionsToPackage{dvipsnames}{xcolor}
\documentclass[aps,pra,twocolumn,amsmath,amssymb,superscriptaddress,floatfix]{revtex4-2}

\usepackage[utf8]{inputenc}
\usepackage[braket, qm]{qcircuit}
\usepackage[ruled,vlined]{algorithm2e}
\usepackage{amsmath}
\usepackage{amssymb}
\usepackage{amsthm}
\usepackage{bbold}
\usepackage{braket}
\usepackage{graphicx}
\usepackage{xcolor}
\usepackage{import}
\usepackage{textcomp}
\usepackage{epstopdf}
\usepackage{booktabs}
\usepackage{mathtools}
\usepackage{subcaption}
\usepackage{caption}
\usepackage{float}

\usepackage{natbib}
\usepackage{wrapfig}

\usepackage[colorlinks=true, citecolor=RawSienna, linkcolor=MidnightBlue, urlcolor=MidnightBlue, linktocpage=true]{hyperref}

\newcommand{\dya}[1]{\ket{#1}\!\bra{#1}}
\newcommand{\defn}{\coloneqq}

\newcommand{\Tr}{{\rm Tr}}

\renewcommand{\vec}[1]{\boldsymbol{#1}}  

\newcommand{\thv}{\vec{\theta}}

\begin{document}

\author{Enrico Fontana}
\email[]{enrico.fontana@npl.co.uk}
\affiliation{Department of Computer and Information Sciences, University of Strathclyde, 26 Richmond Street, Glasgow  G1 1XH, UK}
\affiliation{National Physical Laboratory, Teddington, TW11 0LW, United Kingdom}
\author{Nathan Fitzpatrick}
\affiliation{Cambridge Quantum Computing Ltd, 9a Bridge Street, Cambridge, United Kingdom}
\author{David Mu\~{n}oz Ramo}
\affiliation{Cambridge Quantum Computing Ltd, 9a Bridge Street, Cambridge, United Kingdom}
\author{Ross Duncan}
\affiliation{Cambridge Quantum Computing Ltd, 9a Bridge Street, Cambridge, United Kingdom}
\affiliation{Department of Computer and Information Sciences, University of Strathclyde, 26 Richmond Street, Glasgow G1 1XH, UK}
\affiliation{Department of Physics, University College London, Gower Street, London WC1E 6BT, UK}
\author{Ivan Rungger}
\email[]{ivan.rungger@npl.co.uk}
\affiliation{National Physical Laboratory, Teddington, TW11 0LW, United Kingdom}

\title{Evaluating the noise resilience of variational quantum algorithms}

\begin{abstract}
We simulate the effects of different types of noise in state preparation circuits of variational quantum algorithms.
We first use a variational quantum eigensolver to find the ground state of a Hamiltonian in presence of noise, and adopt two quality measures in addition to the energy, namely fidelity and concurrence. 
We then extend the task to the one of constructing, with a layered quantum circuit ansatz, a set of general random target states.
We determine the optimal circuit depth for different types and levels of noise, and observe that the variational algorithms mitigate the effects of noise by adapting the optimized parameters.
We find that the inclusion of redundant parameterised gates makes the quantum circuits more resilient to noise.
For such overparameterised circuits different sets of parameters can result in the same final state in the noiseless case, which we denote as parameter degeneracy. Numerically, we show that this degeneracy can be lifted in the presence of noise, with some states being significantly more resilient to noise than others. 
We also show that the average deviation from the target state is linear in the noise level, as long as this is small compared to a circuit-dependent threshold. In this region the deviation is well described by a stochastic model. Above the threshold, the optimisation can converge to states with largely different physical properties from the true target state, so that for practical applications it is critical to ensure that noise levels are below this threshold.
\end{abstract}

\maketitle

\pagebreak

\section{Introduction}
The rapid development of noisy intermediate-scale quantum (NISQ) computers \cite{preskill2018quantum} in recent years has seen the equally explosive rise of hybrid variational quantum algorithms (VQAs) \cite{Peruzzo2014, McClean2016, PhysRevA.99.062304, khatri2019quantum, Cerezo2020variationalquantum, cerezo2020variational, cirstoiu2019variational, xu2019variational, zhou2018quantum}.
VQAs are composed of a quantum subroutine embedded into a classical optimisation loop \cite{Peruzzo2014, McClean2016, Yuan2019}. The quantum side of the algorithm consists of a quantum state preparation stage (or ansatz) with externally controllable parameters, and a measurement stage that returns the value of a cost function from the prepared quantum state.
The classical loop is typically a gradient descent optimiser, which updates the parameters in order to minimise the cost function.

VQAs hold much promise for immediate application to NISQ era devices, not just because they do not require large qubit counts to be useful, but also because they are expected to offer some resilience to the noise that characterizes these devices \cite{Moll2018, sharma2019noise, Colless2018, McClean2017, Yuan2019, PhysRevX.6.031045, li2017hybrid}. 
It has been shown that VQAs can automatically compensate for coherent errors, such as over-/under-rotations \cite{McClean2016, Colless2018, PhysRevX.6.031007}. This stems directly from the variational nature of VQAs, as errors that only shift the position of the cost function minimum do not affect the outcome of the optimisation \cite{McClean2016}. 
The resilience of VQAs to decoherent noise has proved more difficult to characterize, although there is some theoretical evidence that they offer partial resilience to more general stochastic errors \cite{Colless2018, McClean2017}. Variational compiling has been shown to be robust against a noise model containing gate and readout noise \cite{sharma2019noise}. There further exist several methods for error mitigation, which allow to remove some unphysical states \cite{McClean2017, mcardle2018quantum, PhysRevLett.119.180509, mcardle2019error, PhysRevX.7.021050, Endo2018}.
However, it is widely believed that there exists a hard limit to the resilience to decoherent noise of VQAs. Recently it has been shown that noise in the state preparation circuit leads to an asymptotic flattening of the cost function landscape \cite{wang2020noise}. This phenomenon reflects an accumulation of decoherent errors that ultimately makes the optimization untrainable.
The result is valid in the limit of deep quantum circuits, while there remains an intermediate regime in system size and circuit depth, in which VQAs can be run successfully. The understanding of the impact of noise in this intermediate regime remains lacking, and is the question that we address in this paper.
We evaluate quantitatively how specific models of quantum noise impact the performance of simulated VQAs, and to what extent this can be mitigated by the circuit design and optimisation strategy.

The paper is divided into two main parts: the simulation of a variational quantum eigensolver (VQE) and a more general variational algorithm that aims at maximising fidelity with a random target state. For the former, we study the effect of varying the circuits and in particular the effect of adding redundant parameterised gates in presence of noise. As measures of quality we track the energy, fidelity and entanglement of the prepared state.
For the latter, we employ a hardware-efficient ansatz \cite{Kandala2017}, where the circuit is composed of a layer repeated several times. We present numerical results without and with noise-aware parameter optimisation, for various noise levels and number of layers in the circuit. 
We find that VQAs can partly mitigate for quantum noise in the state preparation circuit by optimizing the parameters in presence of noise, and demonstrate that the inclusion of redundant parameters, which in the noiseless case results in degenerate sets of parameters with equal energy, can lead to states with higher resilience to noise. 

Finally, we find that for low noise levels the noise propagation in a VQA circuit targeting a random state is well described by a stochastic model. In this regime, the noisy state is a linearly modified form of the exact state. 
However, in some situations there exists a circuit-dependent noise threshold, above which the optimisation in presence of noise can lead to different states with potentially very different physical properties, such as their entanglement. For practical applications it is important to characterise this threshold in order to ensure that the noise is within the linear regime.

\section{Methods}

\subsection{Variational quantum algorithms}

Mathematically, the ansatz of a VQA can be seen as a parameterised unitary operator $U(\thv)$, with the vector $\thv$ representing the used parameters, that is applied to a fixed initial state $\rho_0$ to yield a desired output state $\rho(\thv) = U(\thv)\rho_0 U(\thv)^{\dagger}$. Here we use density operators $\rho$ to represent quantum states, as they can represent both pure and mixed states. Equivalently, we can associate to the unitary a superoperator $\mathcal{U}(\thv) \rho \defn U(\thv)\rho U(\thv)^{\dagger}$. 
In terms of superoperators the overall quantum circuit operation can be written as
\begin{equation} \label{eq:VQA1}
    \rho(\thv) = \mathcal{U}(\thv)\,(\rho_0) = \mathcal{U}_L(\thv_L) \mathcal{U}_{L-1}(\thv_{L-1}) \cdots \mathcal{U}_1(\thv_1)\, (\rho_0) ,
\end{equation}
where each unitary is parameterised by a set of parameters.

The cost function, $C(\thv)$, typically corresponds to the expectation value of an operator $O$ for the prepared state. The classical optimiser then attempts to find the minimum of the real-valued cost function
\begin{equation} \label{eq:VQA2}
    C(\thv) = \Tr[O\, \rho(\thv)] .
\end{equation}
Eqs. \eqref{eq:VQA1} and \eqref{eq:VQA2} encapsulate the unitary dynamics of a closed system representing a noiseless quantum computer. In our work we simulate this on a classical machine, which can be achieved by representing the operators in the equation as matrices and performing the operations numerically. 

\subsection{Noise model}\label{sec:noise_model}
A real quantum computer is an open quantum system, as it cannot be perfectly separated from its surroundings, and interactions with the environment cause deviations from unitarity, known as decoherence. 
Therefore, our basic unitary evolution needs to be expanded with a noise model.
We use the broad term quantum channel to describe both unitary and non-unitary dynamics, and when the channel is non-unitary it is termed a noisy quantum channel \cite{Schumacher1996}.

We construct the model by interleaving noiseless operations and noisy quantum channels that aim to replicate decoherent processes in quantum computers. Indicating the effect of noisy quantum channels as $\Lambda_l$, with the integer $l$ indexing the specific channel, Eq. \eqref{eq:VQA1} is modified to
\begin{align}
    \rho(\thv) = \Lambda_L \mathcal{U}_L(\thv_L) \Lambda_{L-1} \cdots \Lambda_1 \mathcal{U}_1(\thv_1)\, (\rho_0).
    \label{eq:rhoNoise}
\end{align}
We use identical noise channels throughout the circuit evaluation. Each of these spans all the qubits in the system, and is defined as the application of an identical one-qubit noise channel to every qubit:
\begin{equation}
\rho=\Lambda\, (\rho_\mathrm{in}) = \left(\bigotimes_{i=1}^N \Lambda^{(i)}\right)\,(\rho_\mathrm{in}),
\end{equation}
where $\Lambda^{(i)}$ is the one-qubit channel acting on the $i^\text{th}$ qubit, $\rho_\mathrm{in}$ is the input state and $N$ is the number of qubits. This is termed a product channel \cite{Zhang2019}. The same approximation is used in Ref. \cite{Bassi2008}, and is valid when qubits are sufficiently separated physically and there is only small cross-talk. Note that for simplicity we have also assumed that the noise is identical on every qubit, however in general the qubits of a real quantum computer have noise characteristics that can differ significantly from one another \cite{Arute2019}. 
In our approach we neglect cross-talk and coherent errors, as our main focus is on decoherent noise channels. Similarly, we do not consider readout noise, even though this is an important feature of real quantum devices\cite{Arute2019}. The justification for this choice is that readout noise is independent from ansatz design, while the main question of this paper is to address the effect of a noisy ansatz itself. We also neglect finite sampling (shot) noise, which allows us to employ exact statevector simulations.

We consider three types of noise channels: amplitude damping, phase damping, and symmetric depolarising channels. 
Amplitude damping relates to the relaxation of the qubit from an excited state to its ground state, while phase damping to the loss of phase, perturbing the off-diagonal elements of the density matrix \cite{Nielsen2010}. Two common metrics of qubit quality, the longitudinal relaxation ($T1$) and dephasing ($T2$) times, can be related directly to amplitude and phase damping \cite{HU2002, Sawaya2016, Rost2020}.
Symmetric depolarising noise describes a decay to a completely mixed state, and hence is useful as a prototypical decoherent channel \cite{Nielsen2010}.

We make use of the Kraus operator formalism to apply these channels onto the quantum state, which carries the premise that quantum noise is a Markovian process \cite{Nielsen2010}.
In our implementation, the operators are parameterised by $\gamma \in [0, 1]$, representing the strength of the noise. A value of 0 represents no noise (and hence an identity channel), while a value of 1 is maximal in the sense that the output of the noise channel corresponds to the fixed point of the channel. For amplitude damping, this is the state $\dya{0}$, for dephasing it is any linear combination of $\dya{0}$ and $\dya{1}$, and for symmetric depolarising it is the completely mixed state. In Appendix \ref{app:kraus} we provide the single-qubit Kraus operators used for the noise model, as well as details on how to obtain the product channel operators. 

As a simplification we apply noise only on two-qubit gates. This is justified by physical considerations valid for most hardware systems, where 2-qubit operations are considerably slower than the 1-qubit ones, and have a much higher noise rate, usually by an order of magnitude in their gate fidelity figures \cite{Tannu2018}. Consistently with previous work \cite{Lilly2020, Bassi2008}, we place a noisy channel after the gate, ensuring that at maximum noise the state output by the circuit will be unentangled even for non-depolarising noise channels like phase and amplitude damping.

\subsection{Quality measures}\label{sec:methods_outline}

We investigate two classes of variational quantum algorithms: VQE, where the energy is minimised, and target state optimisation, where the infidelity with respect to a random target state is minimised. 

\subsubsection{VQE simulations}

As first system we examine the VQE, the prototypical example of VQA \cite{Peruzzo2014}, and the results are presented in Sec.~\ref{sec:hamopt}. The VQE algorithm seeks to identify the ground state of a given Hamiltonian $H$.
The cost function of the VQE is determined by setting $O=H$ in Eq. \eqref{eq:VQA2}, which returns the energy of the trial state:
\begin{equation}
C(\thv) = E(\thv) \defn \text{Tr}\left[H\rho({\thv})\right] .
\end{equation}

Besides the energy of the state, to assess the quality of the output after convergence we also consider the fidelity with respect to the exact ground state of $H$. The fidelity between two general states $\rho$ and $\sigma$ is defined as \cite{Jozsa1994}
\begin{align}
F(\sigma, \rho) &\defn \left(\text{Tr}\sqrt{\sqrt{\rho}\,\sigma\,\sqrt{\rho}}\right)^2\nonumber\\
&= \text{Tr}\left[\sigma \rho\right]\label{eq:fidelity},
\end{align}
where the last equality is valid whenever one of the two states is pure. If we denote the exact ground state as $\ket{\psi_{\text{gs}}}$, the fidelity for a parameterised ansatz circuit corresponds to
\begin{equation}
F({\thv}) \defn \text{Tr}\left[\dya{\psi_{\text{gs}}}\rho({\thv})\right] = \bra{\psi_{\text{gs}}}\rho({\thv})\ket{\psi_{\text{gs}}},
\end{equation}
where $\rho({\thv})$ is obtained with Eq. (\ref{eq:rhoNoise}).

Finally, since in VQE we are almost exclusively interested in Hamiltonians with entangled ground states, we use entanglement as a further important test of the quality of the output state. 
If we have a 2-qubit system, there exists a broad selection of measures of bipartite entanglement. Since the noisy circuit produces mixed states, we choose one that is valid in this regime, namely the concurrence, $Q(\rho)$ \cite{hill1997entanglement}. The concurrence is frequently used in literature, as it is monotonically related to entanglement of formation, a meaningful measure of entanglement, while being easier to calculate in practice \cite{coffman2000distributed,wootters1998entanglement,article}. It has the closed form
\begin{equation}
Q(\rho) \defn \text{max}(0, \lambda_1-\lambda_2-\lambda_3-\lambda_4) .
\end{equation}
Here $\lambda_i$ are the eigenvalues, in decreasing order, of the Hermitian matrix $\sqrt{\sqrt{\rho}\Tilde{\rho}\sqrt{\rho}}$ (note the similarity to Eq. \eqref{eq:fidelity}), where $\Tilde{\rho}$ is the spin-flipped density matrix $(\sigma_y\otimes\sigma_y)\rho^*(\sigma_y\otimes\sigma_y)$, with $*$ indicating complex conjugation. $Q(\rho) = 0$ if and only if $\rho$ is a linear combination of product states, and $Q(\rho) = 1$ if and only if $\rho$ is a Bell state \cite{Wootters2001}.

When working with larger qubit numbers we cannot directly apply concurrence. Out of the several possible measures of multi-qubit entanglement\cite{Love2007}, we choose the maximum concurrence taken over all pairs of qubits in the system \cite{Y_na__2007, Meill2019}.

\subsubsection{Random target state fidelity optimisation}

In the second part of the article we investigate how noise affects a variational algorithm for solving a more general task. The results are presented in Sec. \ref{sec:randomstates}. We consider random state fidelity optimisation, where, rather than choosing a specific Hamiltonian and evaluating its ground state, we select at random a wavefunction as target. We address the question of how closely a layered quantum circuit can approximate such a general target wavefunction under different noise regimes. The optimisation procedure is therefore modified to maximising the fidelity (see Eq. \eqref{eq:fidelity}) with a target state $\rho_{\mathrm{T}}$. Equivalently, the problem can be formulated as a minimisation of the infidelity, defined as 
\begin{equation}\label{eq:infidelity}
    R \defn 1 - F,
\end{equation}
and hence the cost function is
\begin{equation}
C(\thv) = R(\rho_{\mathrm{T}}, \rho(\thv)) \defn 1 - F(\rho_{\mathrm{T}}, \rho(\thv)) .
\end{equation}
We consider pure target states, so that the cost function given in Eq. \eqref{eq:VQA2} can be applied in this case by choosing $O = \mathbb{1} - \rho_{\mathrm{T}}$.

We then extend this to the case where one is provided with a set of $n_\mathrm{T}$ pure target states sampled from a uniform distribution. 
As figure of merit we use the \textit{average optimal infidelity} over the set, which we define as 
\begin{align}
    \bar{R} &=\frac{1}{n_\mathrm{T}} \sum_{n=1}^{n_\mathrm{T}}\underset{\thv}{\text{min }} R(\rho_{\mathrm{T},n},\rho(\thv)) ,
    \label{eq:avg_opt_infidelity}
\end{align}
where $\rho_{\mathrm{T},n}$ is the target state with index $n$. 
The same measure has been used recently in Ref. \cite{Gard2020}.
For each target state, an optimisation procedure is run in presence of noise. $\bar{R}=0$ would imply that the quantum circuit can represent any $N$-qubit state in the ensemble exactly. The addition of noise is expected to increase $\bar{R}$, as mixed states cannot have perfect overlap with pure states.
As the distribution of target states, we consider the Haar distribution over real states, which is the unique uniform distribution over a space of pure quantum states \cite{Dankert2006}. 

We perform two types of numerical simulations: in the first, we optimise the circuit without noise to obtain the optimal ansatz parameters for the ideal case, but evaluate the infidelity using the noisy circuit (``non-reoptimised''); in the second type we start from the noiseless optimum, but then reoptimise the circuit by performing gradient descent with the noise channels in place (``reoptimised'').
The non-reoptimised cost function provides an upper bound to the reoptimised cost function, and the two will be equal only if the location of the minimum is unaffected by noise.

Finally, in order to isolate the effect of noise from other contributions to the infidelity, we consider the average optimal relative infidelity, which we define as
\begin{equation}
    \bar{R}_{\text{rel}} \defn \bar{R} - \bar{R}_{\text{id}},
\end{equation}
where the subscript id indicates the infidelity evaluated in the ideal noiseless case.

\section{VQE for energy minimisation}
\label{sec:hamopt}
In this section we employ VQE for the task of identifying the ground state energy of a fermionic Hamiltonian, first for 2 and then for 4 qubits.

\subsection{Two qubit system}
\label{sec:two_qubits}

\begin{figure}
\centering
\includegraphics[width=0.48\textwidth]{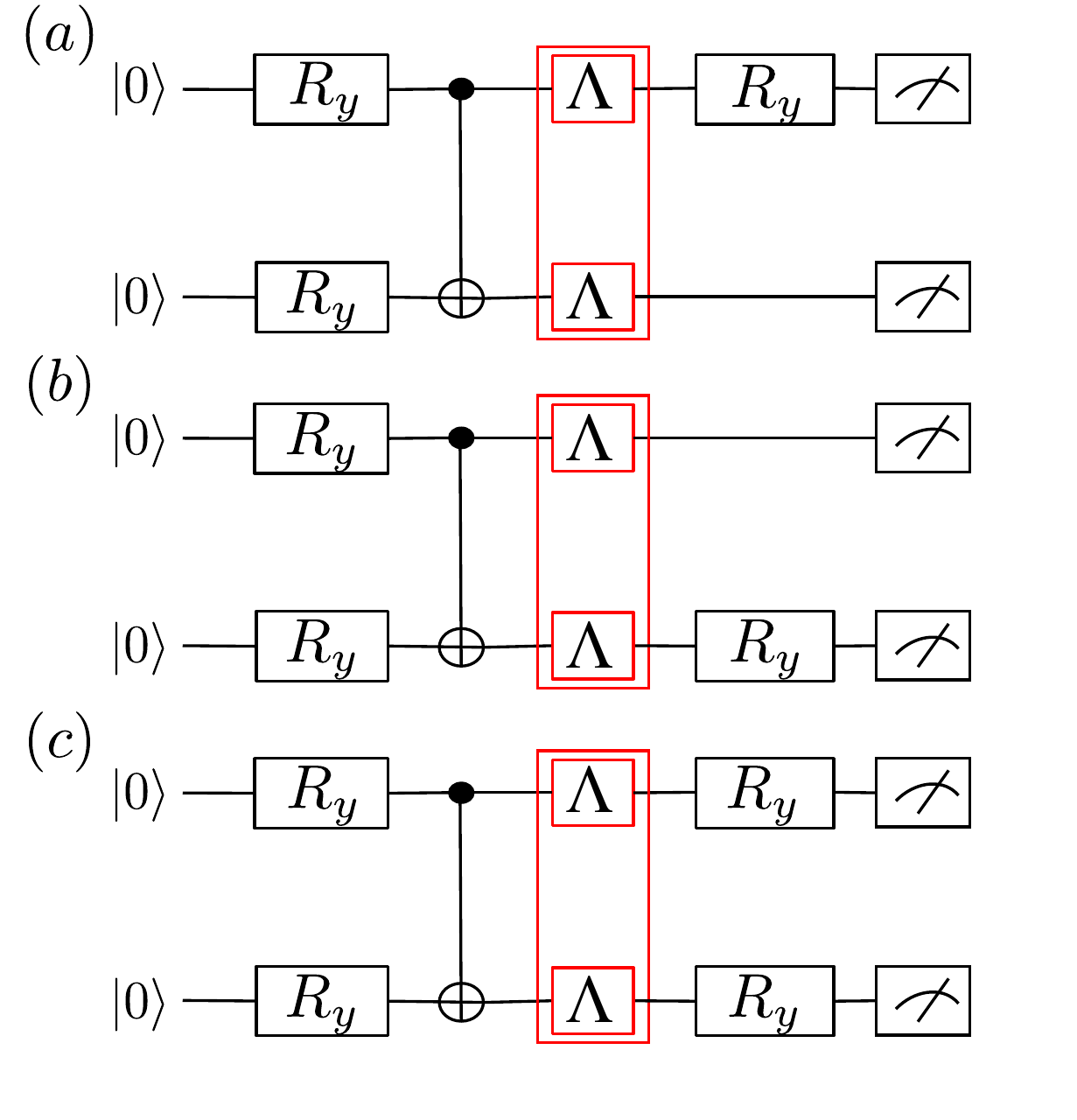}
\caption{Circuit  ans\"atze used for state preparation on two qubits. Circuits (a) and (b) have 3 rotation parameters, and the circuit in (c) has 4 rotation parameters. In the noiseless case they all allow to cover the full real-states space of two qubits, and hence allow to construct any real 2-qubit state. There is one noise channel applied after the CNOT gate, as indicated by the $\Lambda$ blocks on each qubit.
}
\label{fig:circuit_2_qubits}
\end{figure}
We consider the following Hamiltonian on two qubits:
\begin{equation}
\hat{H} = \hat{\sigma}_z^1\hat{\sigma}_z^2 + \hat{\sigma}_x^1 +\hat{\sigma}_x^2 .
\label{eq:hamiltonian}
\end{equation}
This is an example of a transverse-field Ising Hamiltonian \cite{heyl2013dynamical}, and appears in this form in dynamical mean field theory (DMFT) simulations of the single-impurity Anderson model (SIAM) for its 2-electron ground state \cite{rungger2019dynamical}. Since this minimal Hamiltonian has an entangled ground state, it forms an ideal starting point for the investigation of the effect of noise. As will be shown in the subsequent sections, the conclusions found here are applicable also to wavefunctions obtained with more complicated Hamiltonians. Furthermore, as the Hamiltonian is real, we can restrict our choice to just those circuits that always output a real wavefunction, enabling us to significantly reduce the number of parameters. 

\begin{figure*}[t]
\begin{minipage}{0.95\textwidth}
    \centering
    \includegraphics[width=0.95\textwidth]{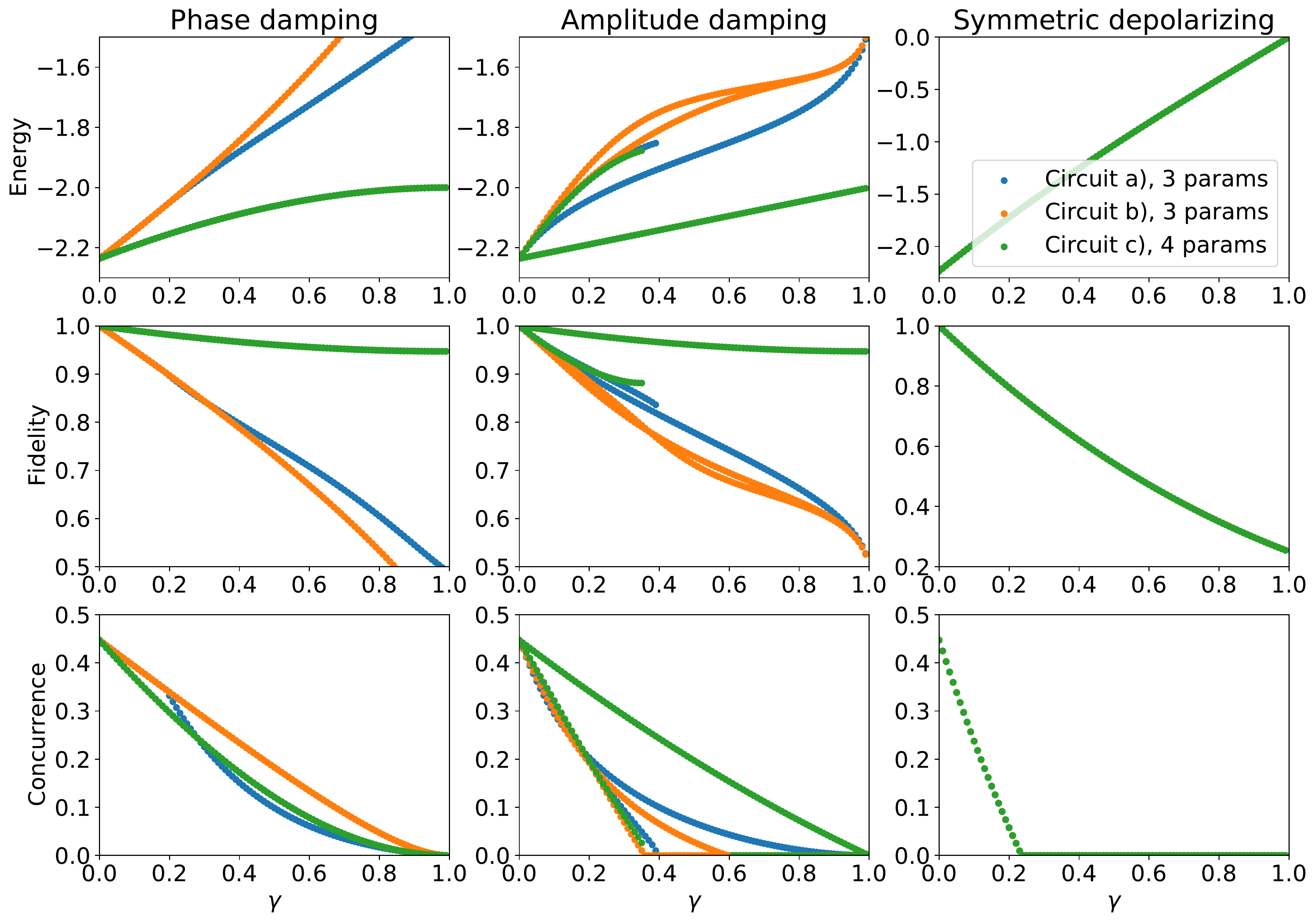}
    \caption{Energy, fidelity and concurrence as function of the noise parameter $\gamma$ for the all energy minima found for the 2-qubit circuits with 3 and 4 gates (Fig. \ref{fig:circuit_2_qubits}), obtained applying the three different indicated noise types. We plot all the obtained local minima in the energy landscape, so that at a given value of $\gamma$ there can be multiple points for the same circuit. For symmetric depolarising noise there is no circuit-dependent difference, and hence only one curve is visible as all curves overlap.
    }
    \label{fig:2_qubits_results}
    \end{minipage}
\end{figure*}
As ans\"atze we choose the three circuits shown in
Fig. \ref{fig:circuit_2_qubits}, each of which can be shown analytically to be able to prepare any possible real 2-qubit state. These ans\"atze include two inequivalent 3-parameter circuits, termed circuit (a) and (b), which differ on the position of the final rotation gate. 
We also consider a 4-parameter circuit that has rotations on both qubits before measurement. This latter circuit is over-parameterised, since it has one extra parameter compared to the previous two, and therefore allows us to explore the impact of redundant parameters.

The VQE algorithm is implemented using an exact density matrix simulator, which allows the use of a gradient-based classical optimiser, specifically the Broyden–Fletcher–Goldfarb–Shanno (BFGS) optimiser in our case \cite{10.1093/imamat/6.3.222, 10.1093/comjnl/13.3.317, 10.2307/2004873, 10.2307/2004840}. The algorithm is evaluated on a range of $\gamma \in [0,1]$, for phase, amplitude damping and symmetric depolarising noise, for all the ansatz circuits.
As outlined in Sec. \ref{sec:noise_model} and Appendix \ref{app:kraus}, $\gamma$ is the parameter of the noise model that encodes the strength of the noise.

Analytically we calculate the exact ground state energy to be $E_{\text{gs}} = -\sqrt{5} \approx -2.236$, and the concurrence to be $Q_{\text{gs}} = \frac{1}{\sqrt{5}} \approx 0.447$. We verified that the numerical simulations with our used ans\"atze reproduce these analytical results exactly for $\gamma = 0$.
The outcome of the noisy VQE simulation is shown in Fig. \ref{fig:2_qubits_results} for all state preparation circuits of Fig. \ref{fig:circuit_2_qubits}. 
Straight away, we notice clear differences between the noise channels. In all measures of state quality, phase and amplitude damping channels are the least destructive, while the symmetric depolarising channel has a much more dramatic effect. 

We perform a comprehensive search of the parameter space, allowing us to identify all the local minima in the energy landscape, which we represent by plotting multiple points for the same value of $\gamma$.
The number of local minima shows a clear dependency on the noise type. Amplitude damping noise reveals local minima that branch out at low noise levels, resulting in multiple points for the same noise value.
The number of minima depends on the circuit: although difficult to deduce from the figure, by numerically analyzing the results we find two solutions for circuits (a) and (b), and three solutions for the 4-parameter circuit.
Symmetric depolarising and phase damping noise instead present a single global energy minimum. 

Importantly, the extent to which different solutions are affected by noise depends strongly on the used circuit. While for symmetric depolarising noise there is no dependence on the circuit, for phase and amplitude damping noise the quality of the prepared state depends strongly on the used circuits. The fact that circuits (a) and (b) give different results shows that in presence of noise shifting a rotation gate from one qubit to another can improve the quality. Furthermore, we consistently see that one of the solutions of the 4-parameter circuit is significantly better than any solution in the 3-parameter circuits, in all measures of state quality. This indicates that for this system the over-parameterised circuit with one redundant angle of rotation exhibits improved capabilities of noise mitigation, or equivalently a higher noise resilience. 

\subsection{Four qubit system}
\begin{figure}
\begin{minipage}{0.45\textwidth}
    \includegraphics[width=\textwidth]{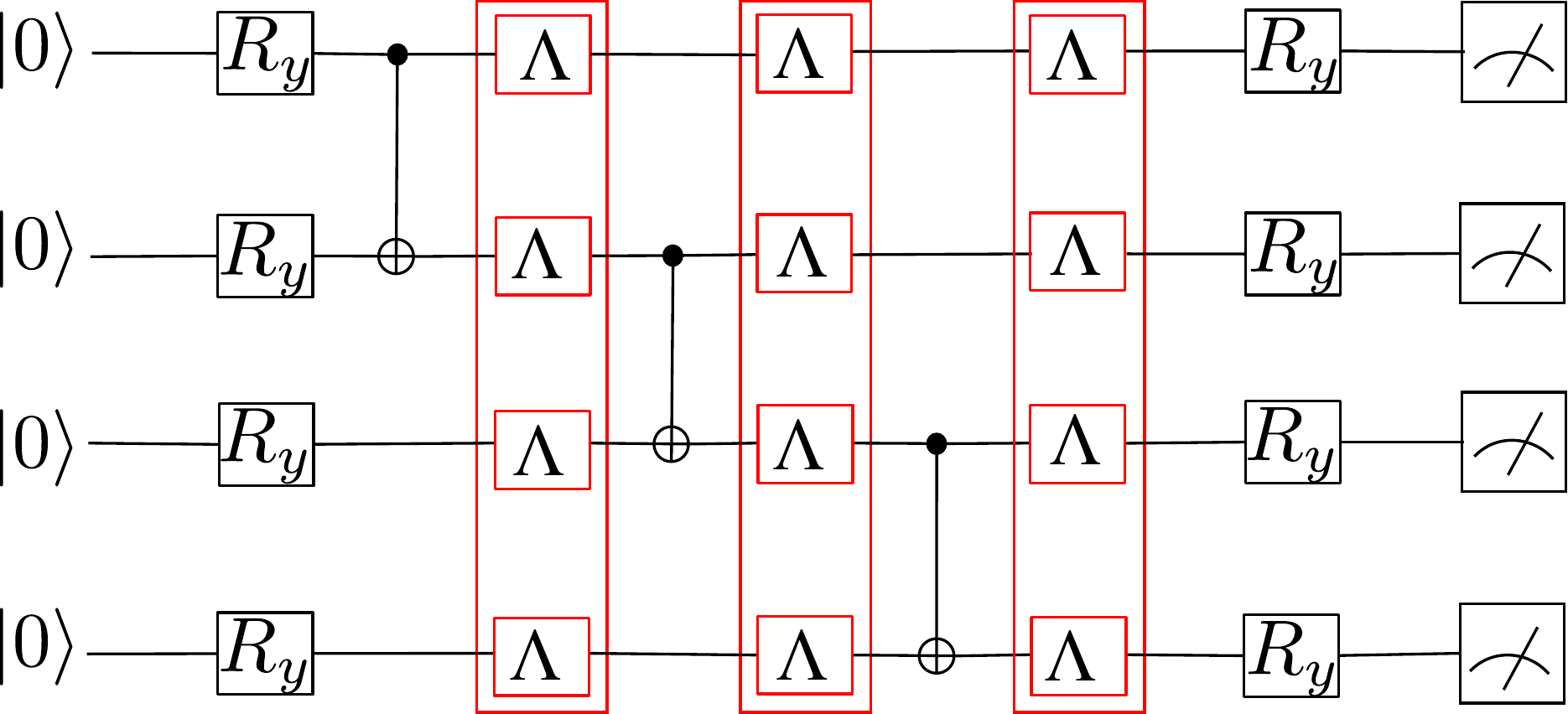}
    \caption{Circuit used for the 4-qubit Hamiltonian VQE simulations. We apply a noise channel on all qubits after each CNOT gate.}
    \label{fig:circuit_4_qubits}
\end{minipage}
\end{figure}

To determine how these findings generalize to higher qubit counts we perform the analogous analysis on a four-qubit system. 
We use the 2-local Hamiltonian
\begin{equation}
\hat{H} = \hat{\sigma}_z^1\hat{\sigma}_z^3 + \frac{1}{2}(\hat{\sigma}_x^1\hat{\sigma}_x^2 + \hat{\sigma}_y^1\hat{\sigma}_y^2 + \hat{\sigma}_x^3\hat{\sigma}_x^4 + \hat{\sigma}_y^3\hat{\sigma}_y^4) .
\end{equation}
This Hamiltonian describes the same physical system as the one used for the previous section, but is valid for any number of electrons in the system \cite{rungger2019dynamical}. For its ground state this Hamiltonian can be projected to the 2-qubit one discussed above, and it follows that the two Hamiltonians have the same ground state energy.
As explained in Section \ref{sec:methods_outline}, since we are dealing with a system larger than 2 qubits we employ the maximum pairwise concurrence as a measure of entanglement. We find that the maximum pairwise concurrence of the new ground state is $Q_{\text{gs}} = \frac{2}{\sqrt{5}} \approx 0.894$.
The chosen state preparation ansatz circuit is shown in Fig. \ref{fig:circuit_4_qubits}, and gives the exact energy in the noiseless case \cite{rungger2019dynamical}.

\begin{figure}
\begin{minipage}{0.48\textwidth}
    \includegraphics[width=\textwidth]{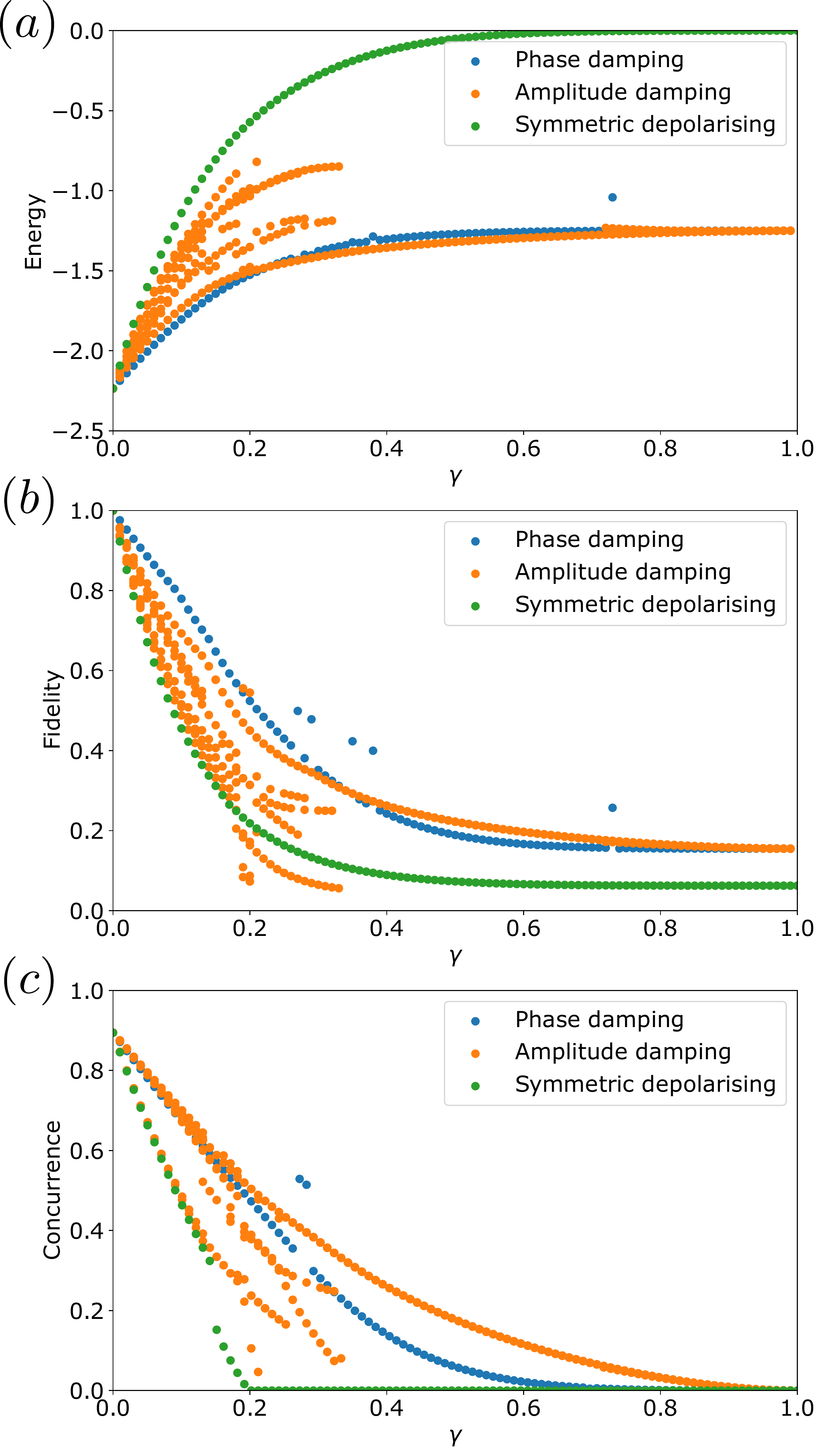}
    \caption{Energy, fidelity and concurrence as function of the noise level $\gamma$ for all the energy minima found for the 4-qubit VQE simulations with the circuit shown in Fig. \ref{fig:circuit_4_qubits}, obtained applying the three different indicated noise types. We plot all the obtained local minima in the energy landscape, so that at a given value of $\gamma$ there can be multiple points for the same circuit. Some points are missing due to imperfect optimisation, and due to the fact that local minima can appear and disappear for increasing noise levels.
    }
    \label{fig:4_qubit_hamiltonian_opt}
\end{minipage}
\end{figure}

The results for maximum concurrence are shown alongside the results for energy and fidelity in Fig. \ref{fig:4_qubit_hamiltonian_opt}. 
We see that the effect of the three types of noise channels on the state optimisation is very similar to the 2-qubit case. In particular, amplitude and phase damping have a less destructive impact than symmetric depolarising.
Furthermore, the 4-qubit system exhibits multiple local minima at low noise for amplitude damping that are absent for phase and symmetric depolarising noise. However, compared to the 2-qubit system, the number of local minima is now larger and their appearance and disappearance more irregular.

\subsection{Discussion}
Overall, the 2-qubit system and the 4-qubit system show a similar behavior with respect to the effect of the three types of noise channels, in all the measures of state quality. Consistently, the most destructive noise channel is the symmetric depolarising channel, where energy, fidelity and concurrence rapidly move away from the exact values. In both simulations the concurrence falls to zero at $\gamma \approx 0.2 - 0.3$.

In contrast, the other two noise channels concede a degree of robustness, yielding better performance even at high noise. Nevertheless, as noise rises towards $\gamma = 1$ all circuits eventually tend towards unentangled states. We note that for this system the concurrence is a much more stringent quality criterion than the energy and fidelity, since it inevitably goes towards zero for all circuits at $\gamma=1$, while in some cases energy and fidelity only deviate by 10-20\% from the exact value at maximum noise. For example, for 2-qubits and complete dephasing, the energy is only about $10\%$ higher than the noiseless value. This shows that the energy alone can be a deceptively poor quality measure for quantum algorithms.

Another common feature of the VQE experiments is the presence of multiple local minima when amplitude damping noise is present in the circuit, which are visualized as multiple lines branching off from the same point at zero noise as the noise parameter is increased. For some of these solutions the measures of quality decrease less with noise, and hence appear to be more resilient than others.

The damping noise models bring to light differences between otherwise equivalent circuits at zero noise. All circuits considered for the 2-qubit case (\ref{fig:circuit_2_qubits}) perform indistinguishably with respect to symmetric depolarising noise. Under the two damping noise channels, however, the three circuits are affected differently, with circuit (c) appearing to perform considerably better than circuits (a) and (b).
Importantly, the different resilience to noise found for circuits (a) and (b) for the 2-qubit case shows that the circuit configuration needs to be optimised for maximisation of noise resilience. We expect that this optimization of gate placement is even more important when the noise level varies across qubits. Our results also show that the inclusion of redundant parameters can further improve the quality of the final state. The origin of this improvement is discussed in the next subsection.

\subsection{Parameter degeneracies}

In this section we show that the noise-induced phenomena discussed above, namely the presence of multiple local minima for amplitude damping and the differing performance of the three circuits with amplitude and phase damping, arise due to general features of parameterised quantum circuits. Let us consider those circuits for which there is a (non-identity) vector map $f$ in parameter space such that:
\begin{equation}
    U(\thv)\ket{\vec{0}} = e^{i \phi}\, U(f(\thv))\ket{\vec{0}},
\end{equation}
indicating equality up to an arbitrary global real phase $\phi$.
If such a map exists, we say that there is a \textit{parameter degeneracy}, since two different sets of parameters result in the same identical state and hence energy (or more generally cost function).
For a given circuit we can have a set of maps of similar form. In case the set is countable, we say that the circuit has \textit{discrete parameter degeneracies}. Conversely, if the members of the set vary continuously across all the set, then the circuit has \textit{continuous parameter degeneracies}.

A circuit with parameter degeneracies will feature symmetries of the cost function in the parameter space, since degenerate sets of parameters must yield the same value of the cost function. 
Therefore, the presence of parameter degeneracies implies the existence of multiple identical minima in the noiseless cost function. In the case of discrete degeneracies the minima are distinct and separated in the cost function landscape, while for a continuous degeneracy the minima are connected and may be visualized as a valley in the landscape.
If one introduces a small state-dependent disturbance in the circuit, which breaks the symmetry between parameter degenerate states, such as for specific noise channels, this leads to different states being produced by the circuit. The degenerate global energy minima for the noiseless case then splits in local minima with different energies.

By analysing the parameters for the equivalent minima for the noiseless case in our 2- and 4-qubit VQE simulations, we find that they obey fixed relations consisting of shifts of the angles by $\pi$ and inversions. We verify that these relations are discrete parameter degeneracies by analytically showing that they preserve the state generated by the circuit. Furthermore, we find numerically that phase damping and symmetric depolarising channels preserve these degeneracies, while amplitude damping channels can break them.
This is thus consistent with the multiple noise induced minima appearing for amplitude damping noise.
In Ref. \cite{DegeneraciesPaper} an algorithm for the systematic construction of a particular set of discretely degenerate parameters is presented.

\begin{figure}
\begin{minipage}{0.5\textwidth}
    \centering
    \includegraphics[width=\textwidth]{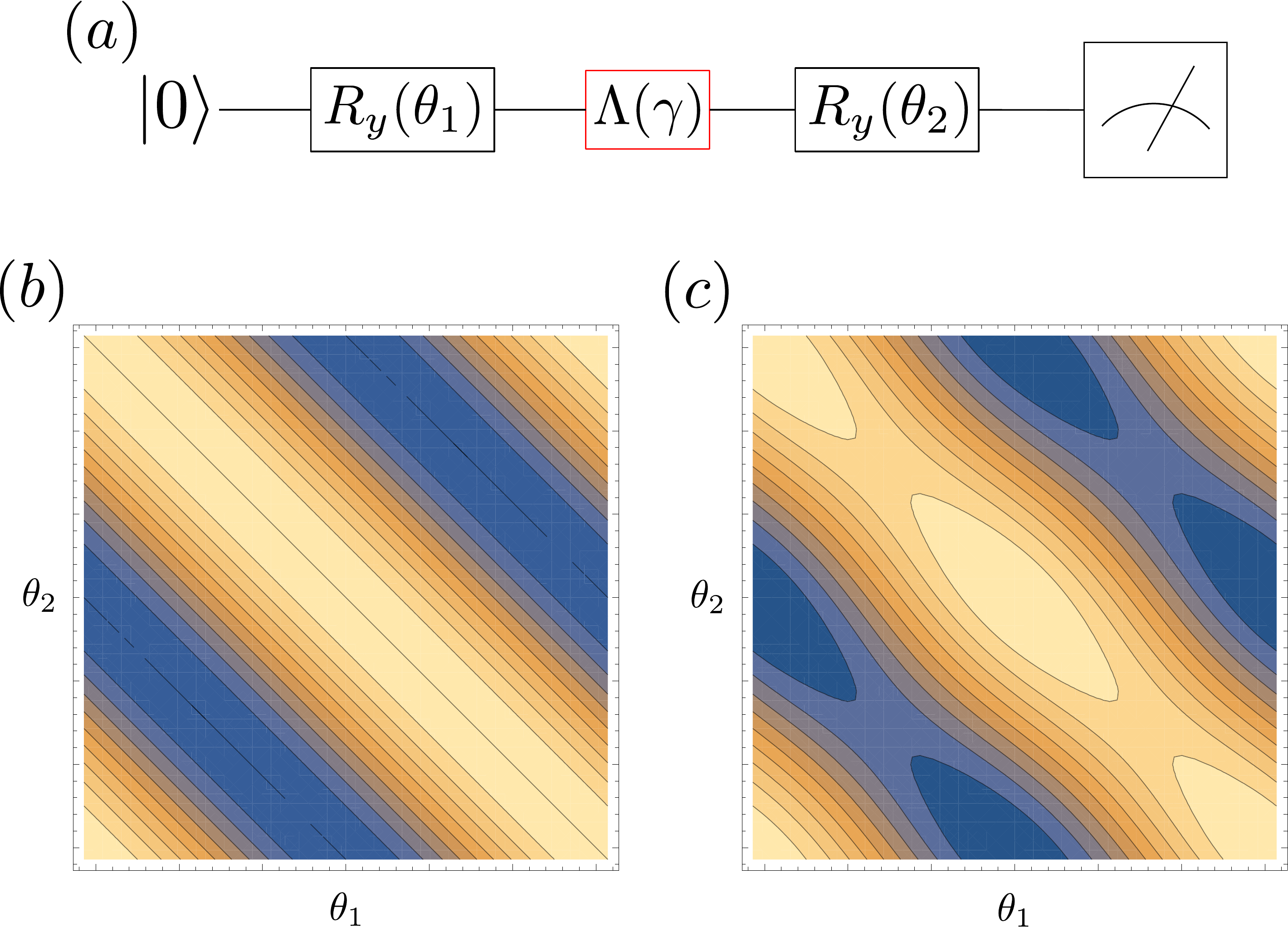}
    \caption{Illustrative example for a continuous parameter degeneracy for an overparametrised circuit, where the second rotation by $\theta_2$ is redundant in the absence of noise; the plots in (a) and (b) show the parameter space landscape of the cost function $C(\theta_1, \theta_2) = \Tr[\rho(\theta_1, \theta_2) \dya{0}]$. In (a) the noiseless results are shown, and in (b) a phase damping channel is added between the Ry gates, with $\gamma = 0.4$. When noise is added, the continuous parameter degeneracy is broken, so that the valley in (a) is replaced by a set of minima in (b).
    }
    \label{fig:broken_degeneracies}
\end{minipage}
\end{figure}

Continuous parameter degeneracies provide a useful framework for explaining the observed improved resilience to noise when including redundant rotation gates. Over-parameterised quantum circuits automatically have continuous parameter degeneracies, since any variation in the redundant parameter can be compensated by modifying the remaining parameters accordingly in order not to change the final state. As illustrative example circuit we consider the placement of two identical single-qubit rotation gates next to each other, shown in Fig. \ref{fig:broken_degeneracies}.
As cost function we consider the overlap with the $\ket{0}$ state: $C(\theta_1, \theta_2) = \Tr[\rho(\theta_1, \theta_2) \dya{0}]$.
The computed cost function exhibits continuous parameter degeneracies in the noiseless case (Fig. \ref{fig:broken_degeneracies}(a)). When a phase damping noise channel is added between the two rotations, the continuous parameter degeneracies are broken, and there now exists a discrete set of minima (Fig. \ref{fig:broken_degeneracies}(b)). For such cases re-optimization of the parameters obtained for $\gamma=0$ for the noisy case will generally lead to improved energies. Since at zero noise the minima exist as a continuous valley, the addition of noise introduces a gradient in this valley, which the optimizer can exploit to arrive to a better solution.
The improved noise resilience for phase and amplitude damping noise due to a redundant parameter can therefore be explained as resulting from the ability of the circuit to explore more state preparation paths in parameter space when compared to a circuit with fewer parameters. It can therefore reach additional paths that are potentially less affected by the noise. For symmetric depolarising noise the relevant parameter degeneracies are not broken upon the addition of noise, and hence the addition of a redundant parameter does not improve the results.

\section{Random target states}
\label{sec:randomstates}
\begin{figure}
\centering
\includegraphics[width=0.45\textwidth]{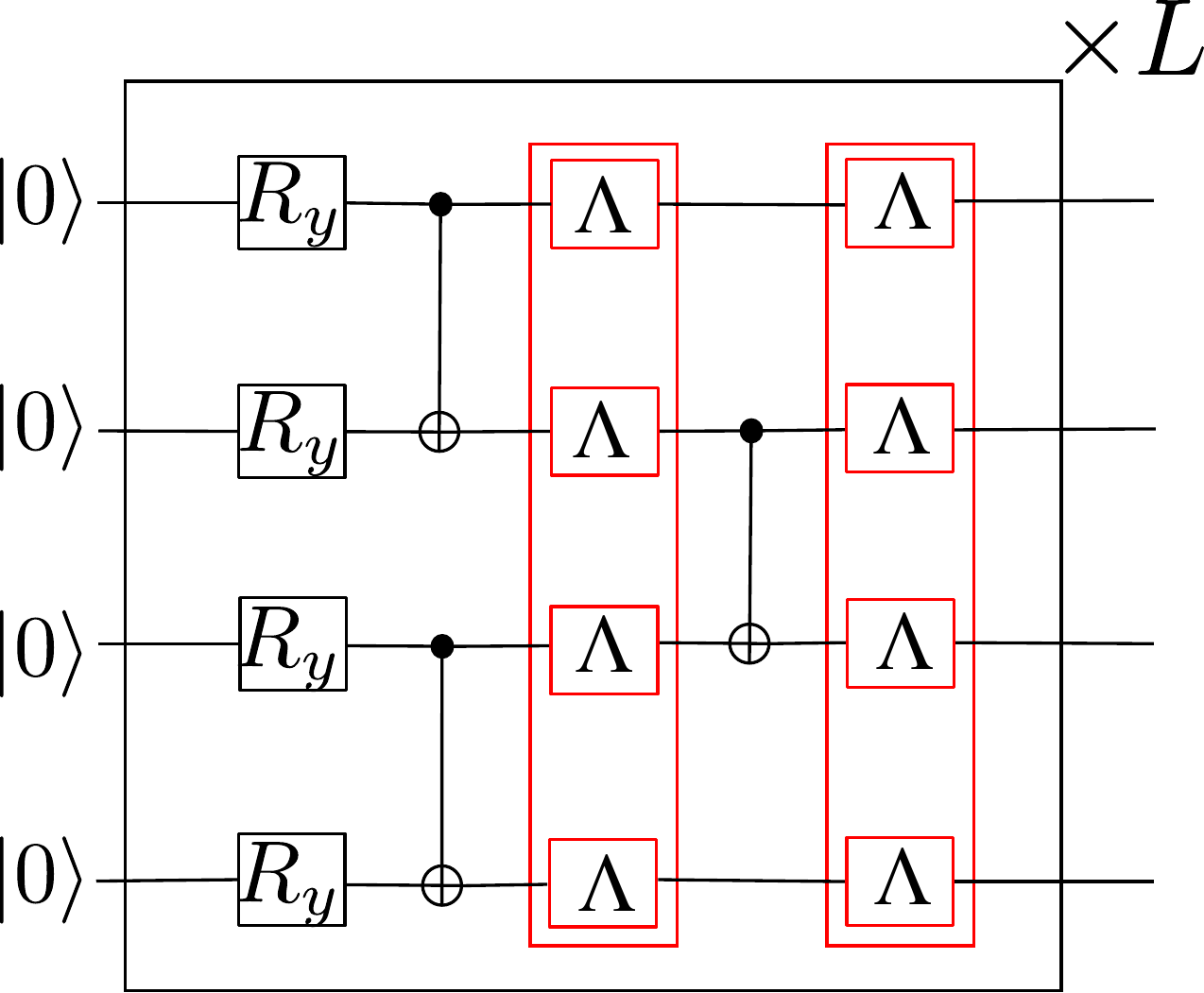}
\caption{Circuit block used for the random target state fidelity maximisation algorithm; shown is one layer of the ansatz, which is repeated a number $L$ times in the full ansatz. Noise channels are added after CNOT gates.
}
\label{fig:optimisation_procedure}
\end{figure}
\begin{figure*}
\begin{minipage}{\textwidth}
    \centering
    \includegraphics[width=1.05\textwidth]{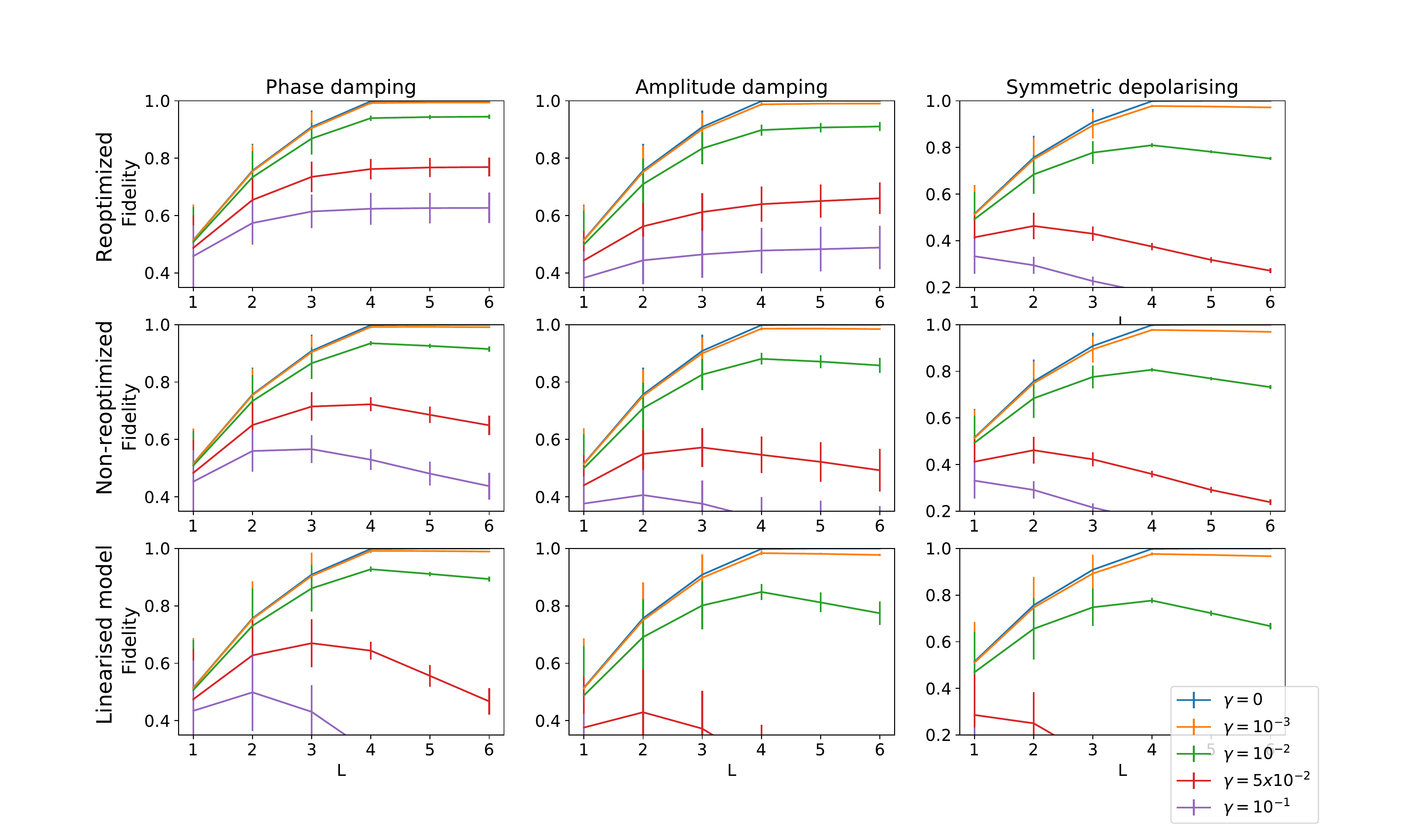}
    \caption{Fidelity vs number of layers, $L$, at realistic noise levels, for the circuit ans\"atze illustrated in Fig. \ref{fig:optimisation_procedure} ($\gamma = 0\%\,\text{(blue)},\, 0.1\%\,\text{(orange)},\, 1\%\,\text{(green)},\, 5\%\,\text{(red)},\, 10\%\,\text{(purple)}$). The plots in the first column are for phase damping noise, the second column plots are for amplitude damping noise, and the third column plots are for symmetric depolarising noise. The top row of plots are for non-reoptimised parameters, in the second row of plots the rotation parameters are reoptimised at each noise level, and the in the third row of plots the linear noise model results are presented. Each point shows the average over 1000 target states, and the vertical bars at each point indicate the standard deviation.
    }
    \label{fig:all_experiments_high_noise}
\end{minipage}
\end{figure*}
To generalise the findings obtained for specific Hamiltonians, here we investigate the ability of a quantum circuit to represent an ensemble of random target states on 4 qubits. This approach allows to estimate how closely ground state wavefunctions of arbitrary Hamiltonians can be reproduced with a given circuit ansatz in presence of noise. For each target state we optimise the circuit to maximise the output state fidelity. We then average the maximum fidelity over the set of random states to obtain the average optimal infidelity (Eq. \eqref{eq:avg_opt_infidelity}). 
We consider again phase damping, amplitude damping and symmetric depolarising noise, and use the same density matrix simulator and local gradient-based minimiser (BFGS method) that we employed in the VQE simulations.
The ensemble considered consists of 1000 real states generated by sampling a random orthogonal matrix from the circular real matrix distribution, the Haar distribution over real orthogonal matrices \cite{Wolfram}, and picking its first column.

We choose a hardware-efficient ansatz \cite{Kandala2017} consisting of an identical layer of 4 rotation gates and 3 CX gates, repeated $L$ times, as illustrated in Fig. \ref{fig:optimisation_procedure}. The choice of this specific structure is motivated by its high expressibility as demonstrated in Ref. \cite{sim2019expressibility}, and by its compactness, which reduces the number of noise channels per layer. Indeed, the first two CX gates can be executed in parallel, and hence according to our noise model we insert only two noise channels per layer.
Hardware-efficient circuits dense in parameterised operations are well-suited for preparing general quantum states \cite{mcardle2018quantum,Romero2019,Moll2018}.

Initially we consider the noise levels $\gamma \in \{0.1\%, 1\%, 5\%, 10\%\}$, which span the range of noise found in current devices \cite{Lilly2020}. The results are shown in Fig. \ref{fig:all_experiments_high_noise}. In the figure we present the average optimal fidelity for each $\gamma$ for the three types of noise channels, where the vertical bars indicate the standard deviation over the ensemble.
For comparison, we also plot the result for the noiseless case ($\gamma=0$).
We consider the two separate cases of noiseless training with noisy evaluation (non-reoptimised), and noisy training with noisy evaluation (reoptimised).
By construction the reoptimised results are always better or equal to the non-reoptimised ones, as the former will take into account any noise-induced change in the cost function landscape. Comparing both results thus gives an insight on the degree to which noise affects the landscape, and on to what extent a variational algorithm can compensate for it. 
From Fig. \ref{fig:all_experiments_high_noise} we can see that the reoptimised results in general improve significantly on the non-reoptimised ones.

We can observe that for $L\geq 4$ the noiseless fidelity is maximised for all the target states and equal to one, with zero standard deviation, showing that  for such overparametrised circuits any target state can be essentially exactly prepared. For $L<4$ on the other hand the circuit does not reach all target states even without noise, and the standard deviation increases as $L$ decreases. 
As expected, the addition of noise further reduces the average fidelity in all plots, with a larger noise level resulting in bigger deviation from the noiseless fidelity. For the non-reoptimised case, with all types of noise the fidelity reaches a peak in $L$, corresponding to the optimal circuit depth in presence of noise. Interestingly, in the case of phase and amplitude damping noise with reoptimisation, the fidelity continues to increase with the number of layers, even for high noise levels.

\begin{figure}[!htbp]
\begin{minipage}{0.45\textwidth}
    \centering
    \includegraphics[width=\textwidth]{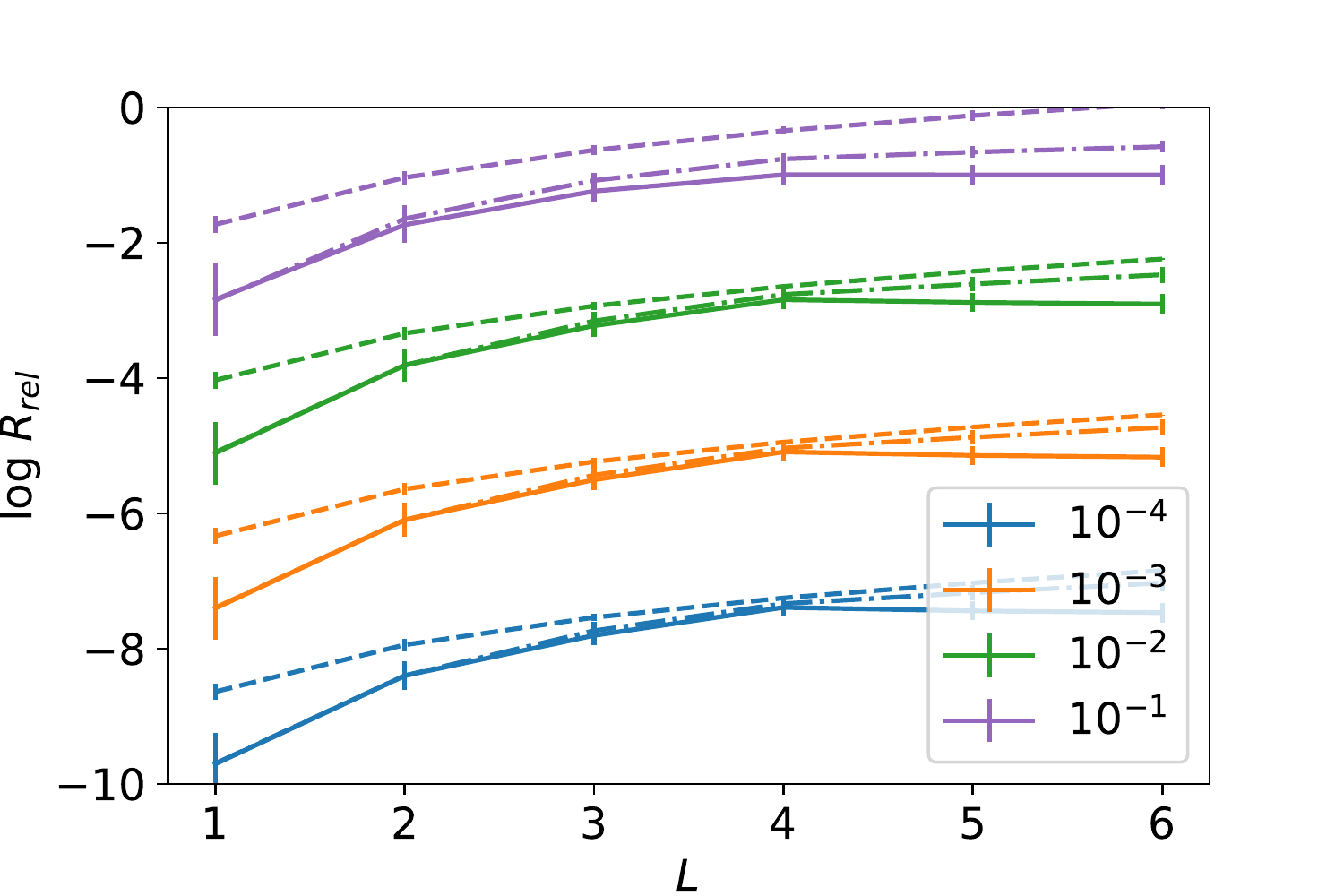}
    \caption{Relative infidelity as function of number of layers, $L$, for different levels $\gamma$ of phase damping noise. The solid curves are for noise-aware reoptimised parameters, the dash-dotted curves are for parameters fixed at their values optimized in absence of noise (non-reoptimised), and the dashed curves indicate the results of the linear noise model. Each point shows the average over 1000 target states, and the vertical bars at each point indicate the standard deviation.
    }
    \label{fig:all_experiments_low_noise}
\end{minipage}
\end{figure}
We also explore noise down to $\gamma = 10^{-4}$, which is representative of the higher quality end of current quantum devices \cite{Lilly2020}. Here we directly compare the relative infidelity as a measure of the effects of noise only, as a function of layers. The results are shown in Fig. \ref{fig:all_experiments_low_noise} for phase damping noise (the results for other types of noise are shown in Appendix \ref{app:plots}).
The relative infidelity reaches a peak at $L=4$ for the reoptimised case, while for the non-reoptimised case it increases monotonically.

\begin{figure}[!htbp]
\begin{minipage}{0.5\textwidth}
    \centering
    \includegraphics[width=\textwidth]{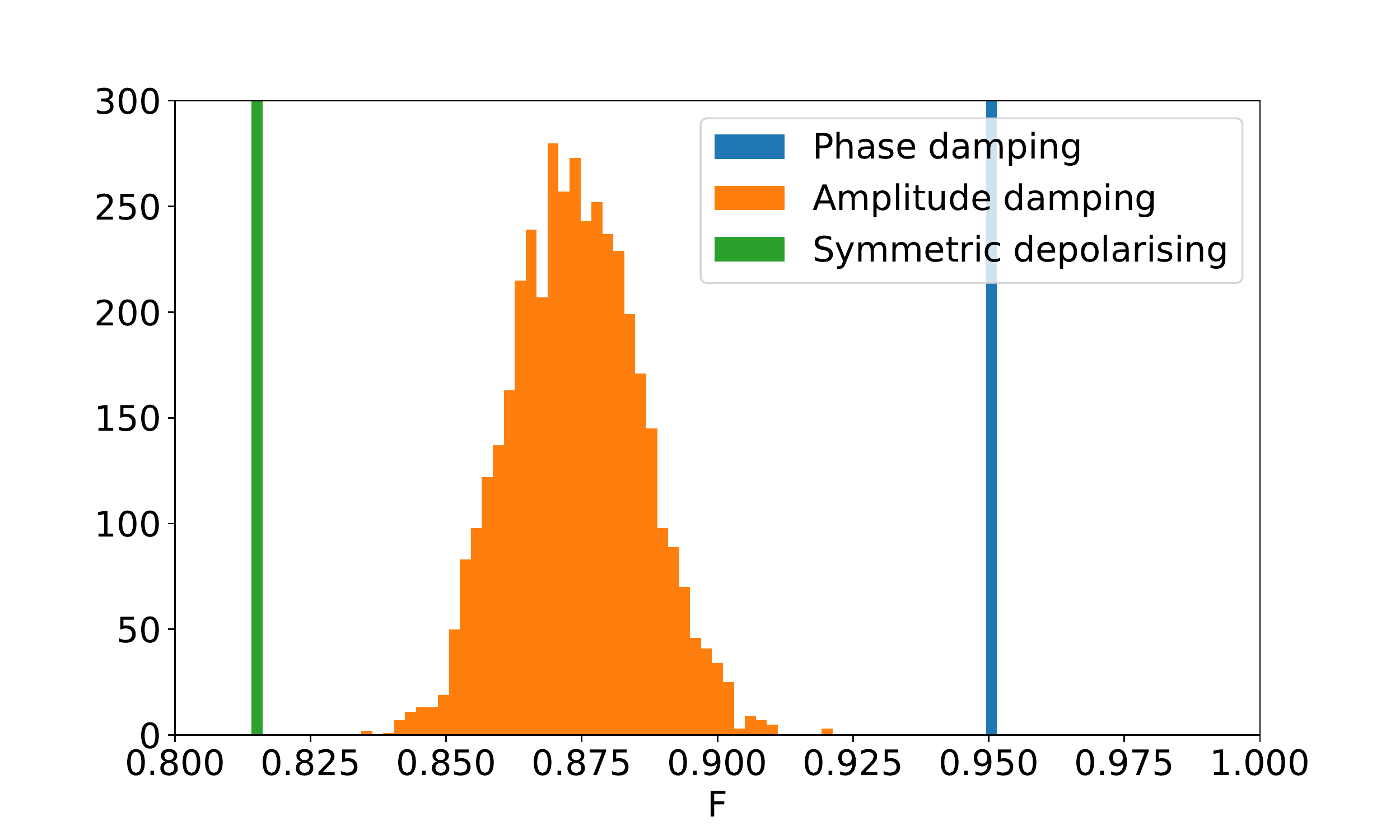}
    \caption{Distribution of fidelity maxima originating from broken parameter degeneracies due to phase damping (blue), amplitude damping (orange), and symmetric depolarising (green) noise. The horizontal axis shows the fidelity, and the vertical axis the number of occurrences in a range of 0.002 around a given fidelity value. The total number of considered discrete parameter degenerate states is 4096, and is constructed using the algorithm described in Ref. \cite{DegeneraciesPaper}. The vertical axis has been cropped for phase damping and symmetric depolarising noise, where all states are within one bin. We choose a random real target state at $L=4$ and $\gamma=0.01$.}
    \label{fig:distribution_degeneracies}
\end{minipage}
\end{figure}

For larger values of $L$ we have an overparametrised ciruit, and we therefore expect the presence of a correspondingly large number of parameter degeneracies. For the 4-qubit VQE system and amplitude damping noise we found a large number of local minima due to discrete parameter degeneracies (Fig. \ref{fig:4_qubit_hamiltonian_opt}). To show this is also the case for general target states we construct a set of degenerate states using the algorithm outlined in Ref. \cite{DegeneraciesPaper}, and evaluate the fidelity for each of these sets of parameters. In Fig. \ref{fig:distribution_degeneracies} we show the resulting distribution of fidelities for $L=4$ as example. While at zero noise they all give the same fidelity, with amplitude damping noise a rather large spread of the fidelities is found. For phase damping and symmetric depolarising noise on the other hand the degeneracies are preserved, and a single sharp peak is found in the histogram. 

\begin{figure}[!htbp]
\begin{minipage}{0.5\textwidth}
    \centering
    \includegraphics[width=\textwidth]{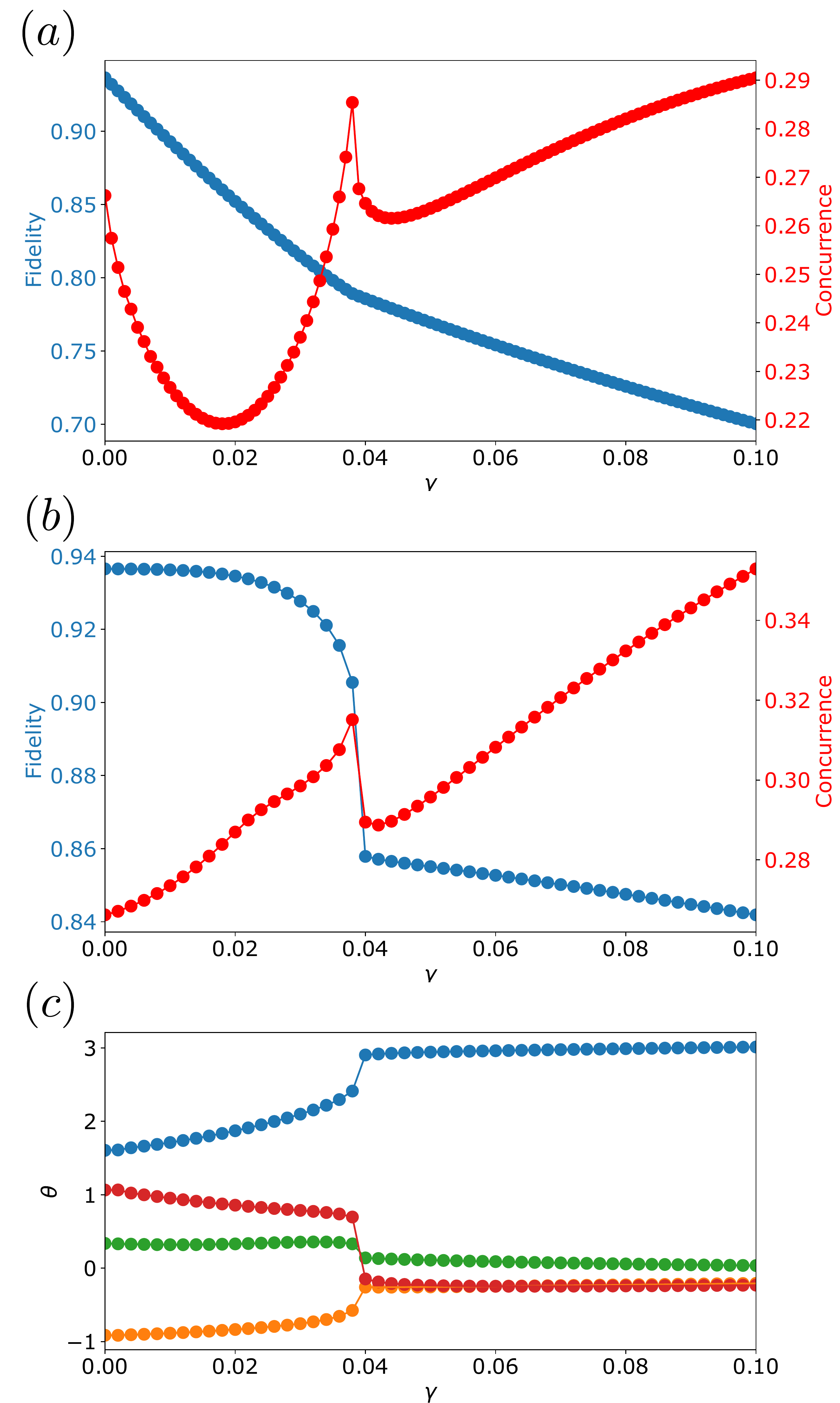}
    \caption{
    Evolution of fidelity and concurrence at convergence for fidelity optimization at different strengths $\gamma$ of phase damping noise, for a single random target state at $L=3$. Four indicative optimised rotation angles are shown in the bottom panel to illustrate the evolution of the optimal parameters with increasing noise. The resulting fidelity and maximum pairwise concurrence are shown in the top panel. The central panel shows the measures evaluated without noise for the angles optimised at each $\gamma$ value. 
    A discontinuity in the slope of both state quality measures and converged angles is found at $\gamma \approx 0.04$, which therefore corresponds to the threshold $\gamma$ value for this state and circuit, above which linear extrapolation of the properties to zero noise is not possible.
    }
    \label{fig:different_minima}
\end{minipage}
\end{figure}

In the VQE optimisation (Section \ref{sec:hamopt}) we observed that under symmetric depolarising and amplitude damping noise, there exists a threshold $\gamma < 1$, past which the algorithm converges to a non-entangled state, corresponding to an undesired noise-induced transition. It is important to verify whether a similar phenomenon appears for the fidelity maximisation with general target states. We therefore choose a target state at random from the distribution, and plot the fidelity of the output state after optimization as a function of noise. The results are shown in Fig. \ref{fig:different_minima} for phase damping noise, $\gamma \in [0, 0.1]$ and $L=3$. In Fig. \ref{fig:different_minima}a we show fidelity and concurrence, and in Fig. \ref{fig:different_minima}c we show a representative subset of the parameters optimized at each $\gamma$. In Fig. \ref{fig:different_minima}b we show fidelity and concurrence evaluated without noise, but with the circuit rotation parameters optimized with noise (Fig. \ref{fig:different_minima}c).
We indeed observe a noise-induced transition in the example considered, which appears at a much lower noise level threshold ($\gamma\approx0.04$) than in the VQE simulations. This transition is also visible in the converged parameter values (Fig. \ref{fig:different_minima}c) and the resulting quality measures evaluated for those parameters without noise (Fig. \ref{fig:different_minima}c).  However, the state after the transition is still entangled, and indeed the concurrence behaves in a nontrivial way, highlighting that the transition is more complex in the general case. We note that the detailed behavior depends on the specific target state, and other examples of such transitions display different behaviours, ranging from sharp thresholds to smoother transitions (see Appendix \ref{app:plots}). There are also target states for which no well-defined transition can be observed.

\subsection{Discussion}

For the non-reoptimised case, the fidelity reaches a peak in $L$, which depends on the noise level, signaling the point where the noise from the increased number of noise channels overcomes the improvement in the accuracy of the circuit with the additional parameters. 
Interestingly, in the case of phase and amplitude damping noise with reoptimisation the fidelity continues to increase with the number of layers even for high noise levels.
The standard deviation of the results shows that the most variability in the state quality is present for shallower circuits, and generally decreases with a larger number of layers. This suggests that, while the accuracy of shallow circuits depends heavily the target state, overparametrised circuits with more layers become increasingly consistent at approximating a general state. Interestingly, the pattern still holds in the presence of noise, however there is additional variance, showing the state-dependence of the noise effects.
Past $L=4$ the reoptimised simulations show a noticeable improvement in noisy fidelity compared to the non-reoptimised simulations. This depth threshold is significant, as it marks the point past which the circuit can perfectly reproduce all target states at zero noise.
Any additional layer beyond $L=4$ therefore does not contribute to the noiseless fidelity and only introduces redundancy. In the noisy case this overparametrisation leads to improved resilience for amplitude and phase damping noise, while for symmetric depolarising noise no improvement is found. This is analogous to what found for the VQE simulations in Sec. \ref{sec:hamopt}.
The specificity to amplitude and phase damping noise is due to the fact that these break the continuous degeneracies resulting from an over-parameterisation of the quantum circuit, while symmetric depolarising noise does not.
Note that we do not expect the improvement to continue to arbitrarily large number of layers, as presumably the capability of over-parametrisation to minimise noise is bounded. More research on real hardware is needed to understand the practical limits.

Improved state quality upon parameter reoptimisation in quantum algorithms has previously been reported in Refs. \cite{Sawaya2016} and \cite{McClean2017}, where the authors find that phase and amplitude damping noise generally impact state preparation in VQE less than symmetric depolarising noise, with phase damping noise being the least impactful. In particular, Ref. \cite{McClean2017} studies how reoptimisation under noise significantly improves the results of a VQE for a chemical problem. However, their model applies noise after the state has been prepared, as opposed to the interleaved noise model proposed here. Hence, the breaking of parameter degeneracies could not be observed in that context.

\subsubsection{Stochastic model}
To provide further insight in our numerical results, here we present a model for the infidelity at small $\gamma$.
In Appendix \ref{app:stochastic_linear_model} we formulate a model that approximates noise propagation as linear, in the sense that each noise channel contributes an additive factor to the final relative infidelity.
The model gives an estimate for the average relative infidelity and its variance as
\begin{align}
    \bar{R}_\mathrm{rel} &\approx \alpha\,\gamma\, d, \\
    \Delta^2_{\text{rel}} &\approx \beta\,\gamma^2 d^2,
\end{align}
where $\alpha$ and $\beta$ are constants obtained from the target state distribution and the noise channel, and $d$ is the number of noise channels, which for our ansatz if $d = 2L$. In Appendix \ref{app:stochastic_linear_model} we provide a detailed description including the procedure for the calculation of $\alpha$ and $\beta$, together with their numerical values obtained for our systems.

We plot the expectations of the model for the fidelity in Fig. \ref{fig:all_experiments_high_noise} (bottom row of panels), and for the relative infidelity in Fig. \ref{fig:all_experiments_low_noise}. Overall the model captures the numerical trends rather well. As expected from the model being a linear approximation, it describes better the behavior of the fidelity at low noise levels, while for higher noise levels the deviations compared to the numerical results become larger. Since the model is formulated under the assumption of no noisy optimisation, it matches the non-reoptimised results much better than the reoptimised ones.
In the latter case, the agreement is nevertheless reasonable until $L<4$, however it diverges significantly for larger $L$, since the model does not include the improvement of fidelity due to parameter reoptimisation with the number of layers for phase and amplitude damping noise. 

A further effect of the noisy optimisation that the linear model cannot capture is the sudden transition to more noise-resilient set of parameters above a noise threshold, as observed in Fig. \ref{fig:different_minima}. Clearly, this poses an upper bound to the noise level that such simplified models can adequately describe, as there exists a threshold noise level past which the new state cannot be extrapolated back to the noiseless state. 
For noise mitigation techniques that extrapolate finite noise data down to the zero noise level\cite{PhysRevLett.119.180509, PhysRevX.7.021050}, this implies that noise data needs to be collected below this critical threshold to avoid extrapolating to potentially undesired zero noise states.\\

\section{Conclusions}
We study the effects of different types and levels of noise on the quality of the results of VQAs. We find symmetric depolarising noise to be the most detrimental, while for amplitude and phase damping noise it is possible to mitigate the effects of noise by optimized gate placement, overparametrisation and noise-aware reoptimization.
We obtain these results consistently across our considered systems, a 2- and 4-qubit simulation for a specific Hamiltonian, and 4-qubit simulations for general target states. 
We introduce the concept of parameter degeneracies, which are sets of parameters in the quantum circuit ansatz that give the same identical output state in the noiseless case. When noise is added, these degeneracies can be broken, leading to some of the originally degenerate parameter states to be more resilient to noise than others.
For the VQE simulations we use three measures of state quality, namely energy, fidelity and entanglement, and show that the energy alone can be a deceptively poor quality measure for quantum algorithms.
When maximising the fidelity of the state produced with a given ansatz with a target state, we find that in presence of symmetric depolarising noise there is a circuit depth, where the fidelity is maximised. For amplitude and phase damping noise, and for the considered circuit depths, noise-aware parameter reoptimisation allows to progressively improve the fidelity as the circuit depth is increased.
The results without noise-aware optimization compare well with a linearised noise model.
We show that the average deviation from the target state is linear for low enough noise levels. For a number of target states and circuits there is a noise threshold, above which the states produced by the circuit can have largely different physical properties from the true target state. For practical applications it is critical to ensure that noise levels are below this threshold in order to preclude convergence to unphysical solutions. 

\section{acknowledgements}
EF and IR acknowledge the support of the UK government department for Business, Energy and Industrial Strategy through the UK national quantum technologies programme. EF acknowledges the support of an industrial CASE (iCASE) studentship, funded by the UK Engineering and Physical Sciences Research Council (grant EP/T517665/1), in collaboration with the universallyersity of Strathclyde, the National Physical Laboratory, and Cambridge Quantum Computing.
We thank Marco Cerezo, Lingling Lao, Dan Browne, Andrew Patterson and Prakash Murali for useful discussions. In part of our simulations we use the software quantumsim \cite{quantumsim}.

\clearpage

\appendix

\vspace{0.5in}

\begin{center}
  {\Large \bf \MakeUppercase{Appendix}}
\end{center}

\section{Kraus operators for simulations of noise}
\label{app:kraus}
In the Kraus operator formalism, each noise channel is assigned a set of operators $\{E_i\}$, which are applied to a quantum state by conjugation. The integer index $i$ spans across all considered operators for a given channel. For a given input quantum state $\rho_\mathrm{in}$ the noisy output state $\rho=\Lambda(\rho_\mathrm{in})$ is obtained by 
\begin{equation}
    \label{eq:kraus}
   \rho= \Lambda (\rho_\mathrm{in})=\sum_k{E_k\rho_\mathrm{in} E^{\dagger}_k}.
\end{equation}
In order to preserve the trace of the quantum state, Kraus operators need to obey the condition 
    $\sum_k{E^{\dagger}_k E_k} = 1$.

The one-qubit phase damping channel is implemented with the following Kraus operator matrices:
\begin{equation}
    E^{(1)}_1=\begin{pmatrix}
            1 & 0 \\
            0 & \sqrt{1-\gamma}
        \end{pmatrix},\hspace{4pt}
    E^{(1)}_2=\begin{pmatrix}
            0 & 0 \\
            0 & \sqrt{\gamma}
        \end{pmatrix}.
\end{equation}
This channel has the effect of suppressing the off-diagonal components of the density matrix, while keeping the diagonal components unchanged.
The one-qubit amplitude damping channel is described by:
\begin{equation}
    E^{(1)}_1=\begin{pmatrix}
            1 & 0 \\
            0 & \sqrt{1-\gamma}
        \end{pmatrix},\hspace{4pt}
    E^{(1)}_2=\begin{pmatrix}
            0 & \sqrt{\gamma} \\
            0 & 0
        \end{pmatrix}.
\end{equation}
Similarly to the phase channel, the off-diagonal components are suppressed. At the same time, however, the diagonal components are altered in favour of the $\ket{0}$ state, representing a relaxation of the system towards the ground state.

Finally, the operators for a single-qubit symmetric depolarising channel are:
\begin{align}
    E^{(1)}_1=\sqrt{1-\frac{3\gamma}{4}}\mathbb{1}_2,& \hspace{4pt}
    E^{(1)}_2=\sqrt{\frac{\gamma}{4}}\sigma_x \nonumber\\
    E^{(1)}_3=\sqrt{\frac{\gamma}{4}}\sigma_y,& \hspace{4pt}
    E^{(1)}_4=\sqrt{\frac{\gamma}{4}}\sigma_z ,
\end{align}
where $\sigma_{x/y/z}$ are the Pauli matrices. Note that we adopted a scaling of the parameter $\gamma$ for symmetric depolarising noise, such that at $\gamma = 1$ the output state is the completely mixed state $\frac{1}{2}\mathbb{I}_2$.
In the Bloch sphere representation of a one-qubit state, pure states are represented as living on the surface of the sphere, while mixed states live in the interior, with the completely mixed state at the center. A symmetric depolarising channel can therefore be visualized as a uniform contraction of the Bloch vectors towards the center. 

For a product channel, the set of Kraus operators is given by the tensor product of the set of single-qubit Kraus operators:
\begin{equation}
    \{E^{(n)}_i\} = \{E^{(1)}_i\}^{\otimes N} .
\end{equation}
For example, the Kraus operators for a local dephasing channel on two qubits are:
\begin{align}
    E^{(2)}_1&=E^{(1)}_1\otimes E^{(1)}_1, \hspace{4pt}
    E^{(2)}_2=E^{(1)}_1\otimes E^{(1)}_2, \nonumber\\
    E^{(2)}_3&=E^{(1)}_2\otimes E^{(1)}_1, \hspace{4pt}
    E^{(2)}_4=E^{(1)}_2\otimes E^{(1)}_2.
\end{align}
To apply a product channel using the formula in Eq. \eqref{eq:kraus} would hence require the summation of exponentially many terms in the number of qubits. In practice, on the quantum simulator this is done by working in the Pauli picture. \cite{quantumsimdoc} Specifically, we can represent a general quantum state as a vector in the Pauli basis, and a general quantum channel as a matrix, also known as Pauli transfer matrix. Therefore, the application of a quantum channel, including noise channels, is reduced to a matrix multiplication between the corresponding Pauli transfer matrix and the vectorised quantum state. \cite{greenbaum2015introduction}

\section{Linear model of noise propagation}
\label{app:stochastic_linear_model}
In the interleaved noise model introduced in Sec. \ref{sec:noise_model} we consider a quantum circuit that we split in $d$ layers. 
In the absence of noise, the unitary operation applied by the circuit can be written as:
\begin{equation}
    U = U_d U_{d-1}\cdots U_2 U_1,
\end{equation}
where $U_i$ is the $i^{\text{th}}$ layer unitary operator. In the noisy case we move to the density matrix picture for mixed states, and use maps between density operators (or superoperators) to represent operations on these mixed states, which are defined by $\mathcal{U}(\rho)=U\rho U^\dagger$. This represents a noiseless quantum channel.
As outlined in the main text, this corresponding noisy quantum channel is given by 
\begin{equation}
    \tilde{\mathcal{U}} = \Lambda_d\mathcal{U}_d...\Lambda_1\mathcal{U}_1 .
    \label{eq:noisy_circuit}
\end{equation}

\subsection{Example: global depolarising noise}
We first consider the case where all the noise channels are given by global symmetric depolarising noise on $N$ qubits, since in this case we can write down an exact model for the noise propagation.
The global symmetric depolarising channel is defined as acting on any state $\rho_\mathrm{in}$ as follows:
\begin{equation}
    \Lambda\,(\rho_\mathrm{in}) = (1 - \gamma)\rho_\mathrm{in} + 2^{-N} \gamma \,\mathbb{1},
\end{equation}
where $\mathbb{1}$ is the identity operator on $N$ qubits (in matrix form it is of dimension $2^N\times2^N$). The effect of the channel is therefore to replace any state with the completely mixed state with probability $\gamma$.

If the circuit under consideration consists of $d$ such channels interleaved with unitary operations, the probability of the state not being affected by any channel is $(1 - \gamma)^{d}$. Any other outcome will lead to the final state being maximally mixed, as the maximally mixed state is invariant under the action of any quantum channel.
The final state of the circuit can therefore be written as
\begin{equation}
    \rho = (1 - \gamma)^{d} \mathcal{U}\,(\rho_\mathrm{in}) + 2^{-N} (1 - (1 - \gamma)^d)\,\mathbb{1},
    \label{eq:utildeTotal}
\end{equation}
where $\mathcal{U}$ is the noiseless circuit unitary map.

We can quantify the difference between the state produced by the noisy circuit, $\rho$, and the state produced without noise, $\rho_\mathrm{id}$, using the infidelity, defined in Eq. \eqref{eq:infidelity}. From now on, the subscript $id$ will universally refer to a quantity evaluated on an ideal, noiseless circuit.
For the global symmetric depolarising noise operator in Eq. (\ref{eq:utildeTotal}) we obtain
\begin{align}
    &R(\rho_\mathrm{id}, \rho) \nonumber\\
    &= 1- \text{Tr}[\rho_\mathrm{id} ((1 - \gamma)^{d} \rho_\mathrm{id} + 2^{-N} (1 - (1 - \gamma)^d)\mathbb{1})] \nonumber\\
    &= 1 - (1 - \gamma)^{d} \text{Tr}[\rho_\mathrm{id}^2] - 2^{-N}(1 - (1 - \gamma)^d) \text{Tr}[\rho_\mathrm{id}] \nonumber\\
    &= \left(1-\left(1 - \gamma\right)^{d}\right) \left(1-2^{-N}\right),
\end{align}
where we use the fact that $\rho_\mathrm{id}$ is a pure state, and so $\text{Tr}[\rho_\mathrm{id}^2] = 1$.
Expanding at first order in $\gamma$ we get 
\begin{equation}
    R(\rho_\mathrm{id}, \rho) = (1 - 2^{-N})\,d\,  \gamma + O(\gamma^2). 
\label{eq:fidelity_symmdep}
\end{equation}
This means that, in the low noise regime, the infidelity increases approximately linearly in both $\gamma$ and the number of layers $d$, with a prefactor that is effectively 1 for large systems ($N \rightarrow \infty$).

Note that a global symmetric depolarising channel is not equivalent a product of single-qubit depolarising channels, and therefore the relation derived in Eq. (\ref{eq:fidelity_symmdep}) does not apply for the product channel noise model that we employ in this work.
However, the simple result of Eq. (\ref{eq:fidelity_symmdep}) provides the motivation to seek an analogous equation for more general noise models.

\subsection{Infidelity propagation for general noise channels}
We start by expanding the action of a general $N$-qubit noise channel on a state $\rho$ to first order as
\begin{equation}
    \Lambda\,(\rho) = (1-\gamma)\rho + \gamma\lambda(\rho) + O(\gamma^2) .
    \label{eq:linear_action}
\end{equation}
Here $\lambda$ is a map that represents the first-order action of the channel, which for global symmetric depolarising channels is simply $\lambda(\rho) = 2^{-N} \mathbb{1} \hspace{5pt} \forall \rho$. Note that the matrix $\lambda(\rho)$ is, in general, not a valid quantum state.
We term $\lambda$ the \textit{linear action} of the noise channel.
In the expansion we take into account the fact that for a general noise channel there could be higher order effects in the local noise parameter $\gamma$.
Since the gate noise level in current quantum computers has $\gamma \ll 1$ \cite{Lilly2020}, we expect the linear action to always dominate, and hence that one can neglect higher order terms.

We introduce the partially noisy circuit:
\begin{equation}
    \tilde{\mathcal{U}}_{:i:} := \mathcal{U}_d\cdots\mathcal{U}_{i+1}\lambda_i\mathcal{U}_i\cdots\mathcal{U}_1,
\end{equation}
obtained by inserting a single noise channel in the noiseless circuit and keeping only the terms at order $\gamma$. Applying the noisy circuit onto an initial state $\rho_0$ and expanding in $O(\gamma)$, it reduces to:
\begin{align}
    \rho &\approx [(1-\gamma d)\mathcal{U} + \gamma\sum^d_{i=1}\tilde{\mathcal{U}}_{:i:}]\,(\rho_0)\nonumber \\
  &= \rho - \gamma\sum^d_{i=1}(\mathcal{U}-\tilde{\mathcal{U}}_{:i:})\,(\rho_0).
\end{align}
The resulting difference to the noiseless result is an additive contribution proportional to $\gamma$.

Now we combine the first-order expansion with the layered noise model, and apply it to the variational problem of maximising the fidelity to a pure target state $\rho_{\mathrm{T}}$. In this case, the fidelity between the noisy state output by the circuit after optimisation $\rho_{\text{opt}}$ and the target state simplifies to $\text{Tr}(\rho_{\mathrm{T}}\rho_{\text{opt}})$.
In general, the circuit may not reproduce the ideal target state $\rho_{\mathrm{T}}$ exactly. We therefore have
\begin{align}
    &F(\rho_{\mathrm{T}}, \rho_{\text{opt}}) = \text{Tr}(\rho_{\mathrm{T}}\rho_{\text{opt}}) \nonumber\\
    &\approx \text{Tr}[\rho_{\mathrm{T}}\rho_{\text{opt,id}} - \rho_{\mathrm{T}}\gamma\sum^d_{i=1}(\mathcal{U}-\tilde{\mathcal{U}}_{:i:})(\rho_0)] \nonumber\\
    &= F(\rho_{\mathrm{T}}, \rho_{\text{opt,id}}) - \gamma\sum^d_{i=1}(F(\rho_{\mathrm{T}}, \rho_{\text{opt,id}})-F(\rho_{\mathrm{T}}, \tilde{\mathcal{U}}_{:i:}(\rho_0))) .
    \label{eq:fidelity_propagation}
\end{align}
We can now define the \textit{relative infidelity} as 
\begin{equation}
    R_{\text{rel}}(\rho_{\mathrm{T}};\rho_{\text{opt,id}},\rho_{\text{opt}}) \defn F(\rho_{\mathrm{T}}, \rho_{\text{opt,id}}) - F(\rho_{\mathrm{T}},\rho_{\text{opt}}) .
\end{equation}
We now define the state at the $i$th layer $\rho_{\text{opt,}i}$, as well as the target state "back-propagated" to $i$th layer as
\begin{equation}
    \rho_{\mathrm{T,}i} \defn \mathcal{U}_{:i}^{-1} \rho_{\mathrm{T}},
\end{equation}
where we use the fact that $\mathcal{U}_{:i}$ is a unitary channel. With this definition and the invariance of fidelity under unitary transformations, we can write Eq. \eqref{eq:fidelity_propagation} as:
\begin{align}
  R_{\text{rel}}(\rho_{\mathrm{T}};\rho_{\text{opt,id}},\rho_{\text{opt}}) &\approx \gamma\sum^d_{i=1}R_{\text{rel}}(\rho_{\mathrm{T,}i}; \rho_{\text{opt,}i}, \lambda(\rho_{\text{opt,}i})) .
  \label{eq:model_rel_infidelity}
\end{align}
This result shows that the relative infidelity is approximately linear in the number of noise channels $d$ and the local noise parameter $\gamma$. Each noise channel contributes a factor that depends on the target state $\rho_{\mathrm{T}}$, the state through the channel $\rho_{\text{opt,}i}$ and its image under the linear action $\lambda(\rho_{\text{opt,}i})$. Note that this is consistent with the result for the symmetric depolarising channel in the previous section, by taking
\begin{equation}
    R_{\mathrm{rel}}(\rho_i; \rho_i, \lambda(\rho_i)) = R(\rho_i, \lambda(\rho_i)) = 1 - 2^{-n} \hspace{5pt} \forall \rho_i .
\end{equation}
Importantly, our total infidelity is state dependent. Hence, the path through the space of states that the circuit takes is very important for the noise resilience of the algorithm. If two different paths lead to the same state in the noiseless case, they may still have very different infidelities in the noisy case, as dictated by these intermediate states.

To simplify the result and compare it to literature, one can estimate the average infidelity of a noise channel over all states, $R_i$, and with it obtain a very rough estimate for the total infidelity.
First let us assume that the circuit can indeed reach the target state, such that we have 
\begin{equation}
    R_{\text{rel}}(\rho_{,i}; \rho_{i}, \lambda(\rho_{i})) = R(\rho_{i}, \lambda(\rho_{i})) = 1 - \Tr[\rho_{i}\, \lambda(\rho_{i})]
\end{equation}
and
\begin{equation}
    R_{\text{rel}}(\rho_{\text{id}}; \rho_{\text{id}}, \rho) = R(\rho_{\text{id}}, \rho) = 1 - \Tr[\rho_{\text{id}}\, \rho].
\end{equation}
Now use the definition of the linear action in Eq. \eqref{eq:linear_action} to write, to first order in $\gamma$,
\begin{align}
    \gamma R_{\text{rel}}(\rho_{i}; \rho_{i}, \lambda(\rho_{i})) &\approx \gamma - \Tr[\rho_{i}\, \Lambda(\rho_{i})] + (1 - \gamma)\Tr[\rho_{i}^2] \nonumber \\
    &= 1 - \Tr[\rho_{i}\, \Lambda(\rho_{i})].
\end{align}
In accordance with literature, we define the gate infidelity as $R_i = 1 - \Tr[\rho_{i}\, \Lambda(\rho_{i})]$.
In this way with Eq. (\ref{eq:model_rel_infidelity}) one obtains for the total infidelity of the circuit
\begin{align}
    R \approx \sum^d_{i=1} R_i.
\end{align}
From this equation we obtain the total state-averaged circuit fidelity as function of the averaged fidelities of each noise channel, $F_i=1-R_i$, as
\begin{align}
    F&= 1-\sum^d_{i=1} R_i
    =\prod_{i=1}^d \left(1-R_i\right)+O\left(R_i^2\right).
\end{align}
Since the infidelities of a noise channels are small, we can neglect higher order terms, and obtain 
\begin{align}
    F&\approx \prod_{i=1}^d F_i,
\end{align}
where we have used $F_i=1-R_i$. This last expression is widely used in literature. For example, in Ref. \cite{Arute2019} the fidelity is assumed to behave in this way, and subsequent experimental results are shown to be consistent with this assumption.
In Ref. \cite{Zhou2020} it is shown that the approximation error of a matrix product state representation of a quantum circuit is approximately multiplicative in the fidelity of a single layer. In Ref. \cite{marshall2020characterizing}, a similar reasoning shows that this multiplicative decay of fidelity applies for interleaved depolarising channels in QAOA circuits. The decay of the cost function with the number of noisy channel that follows from these equations was rigorously proven in Ref. \cite{wang2020noise} for local symmetric depolarising channels.
Ref. \cite{Carignan-Dugas2019} shows that infidelity grows at most quadratically with the number of layers of a circuit, and for decoherent channels it is expected to grow linearly at first order, which is equivalent to what we have shown here. However, the measure used there is the gate infidelity over Haar-distributed states, which might not reflect the outcome of specific experiments \cite{sanders2015bounding}.

\subsection{Expected fidelity over an ensemble}
Up to now we considered the infidelity for one target state. We extend this to the case where one is provided with an ensemble of $n_\mathrm{T}$ pure target states sampled from a distribution. 
As figure of merit we use the average optimal infidelity over the ensemble, which we defined in Eq. \eqref{eq:avg_opt_infidelity} as 
\begin{align}
    \bar{R} &=
    \langle R(\rho_{\mathrm{T}},\rho_{\text{opt}})\rangle_\mathrm{T},
\end{align}
where $\rho_{\text{opt}}$ is optimized with noise for each target state.
An $\bar{R}$ of zero would imply that the quantum circuit can represent any $N$-qubit state exactly. In practice $\bar{R}$ is usually larger than zero even in the noiseless case due the inherent limitations of a given circuit, and the addition of noise further increases $\bar{R}$.
We define the average optimal relative infidelity as
\begin{equation}
    \bar{R}_{\text{rel}} \defn \bar{R} - \bar{R}_\mathrm{id} = \langle R_{\mathrm{rel}}(\rho_{\mathrm{T}};\rho_{\text{opt,id}}, \rho_{\text{opt}})\rangle_\mathrm{T},
\end{equation}
where as before the subscript id indicates the infidelity and density matrix evaluated without noise:
\begin{equation}
    \bar{R}_\mathrm{id}=\langle R(\rho_{\mathrm{T}}, \rho_{\text{opt,id}})\rangle_\mathrm{T} .
\end{equation}

Next we derive an approximate relation for $\bar{R}_\mathrm{rel}$ allowing us to qualitatively understand its behavior and relate it to existing literature.
Using Eq. \eqref{eq:model_rel_infidelity} and the fact that in our noise model all the noise channels are identical, we obtain
\begin{align}
  \bar{R}_{\text{rel}} &\approx \gamma \left\langle \sum^d_{i=1} R_{\text{rel}}(\rho_{\mathrm{T,}i}; \rho_{\text{opt,}i}, \lambda(\rho_{\text{opt,}i}))\right\rangle_{\mathrm{T}} .
\end{align}
We now assume that at every level the noiseless $\rho_{\mathrm{T,}i}$ and $\rho_{\text{opt,}i}$ are distributed identically to the final states $\rho_{\mathrm{T}}$ and $\rho_{\text{opt}}$.
Thus we can write
\begin{align}
    \bar{R}_{\text{rel}} &\approx \langle R_{\text{rel}}(\rho_{\mathrm{T}}; \rho_{\text{opt}}, \lambda(\rho_{\text{opt}}))\rangle_{\mathrm{T}}\,\gamma\, d \nonumber \\
                &\approx \alpha\,\gamma\, d ,
    \label{eq:model_infidelity}
\end{align}
where we defined the constant of proportionality
\begin{equation}
   \alpha \defn \langle R_{\text{rel}}\left(\rho_{\mathrm{T}};\rho_{\text{opt}},\lambda(\rho_{\text{opt}})\right)\rangle_\mathrm{T}.
    \label{eq:alpha}
\end{equation}
Note that since $\lambda(\rho_{\text{opt}})$ is generally not a valid quantum state by itself, since usually $\text{Tr}[\lambda(\rho)] \neq 1$, $\alpha$ may also be greater than 1.

With the additional assumption about the distributions of the intermediate states, the variance of the relative infidelity can be estimated. Defining the variance of the optimal relative infidelity over the target states:
\begin{align}
    \Delta^2_{\mathrm{rel}} &\defn \langle R_{\text{rel}}^2(\rho_{\mathrm{T}}; \rho_{\text{opt,id}}, \rho_{\text{opt}})\rangle_{\mathrm{T}} - \bar{R}_{\text{rel}}^2 \nonumber\\
    &= \underset{\mathrm{T}}{\text{Var}} \left(R_{\text{rel}}(\rho_{\mathrm{T}}; \rho_{\text{opt,id}}, \rho_{\text{opt}}) \right) .
\end{align}
Using Eq. \eqref{eq:model_rel_infidelity} and a standard property of the variance, within these approximations we obtain
\begin{align}
    \Delta^2_{\mathrm{rel}} &\approx \underset{\mathrm{T}}{\text{Var}} \left(\gamma\sum^d_{i=1} R_{\text{rel}}(\rho_{\mathrm{T,}i}; \rho_{\text{opt,}i}, \lambda(\rho_{\text{opt,}i})) \right) \nonumber\\
    &= \gamma^2 \underset{\mathrm{T}}{\text{Var}} \left(\sum^d_{i=1} R_{\text{rel}}(\rho_{\mathrm{T,}i}; \rho_{\text{opt,}i}, \lambda(\rho_{\text{opt,}i})) \right) .
\end{align}
Now we must introduce some information about the correlation between states at every noise channel. Since states going through successive channels have a certain degree of similarity, in the sense that they are related by a short-depth sequence of unitaries, it should be expected that they retain a high level of correlation. Therefore, we assume that the states at every noise channel $\{\rho_{\text{opt,1}}, \rho_{\text{opt,2}}, \cdots, \rho_{\text{opt,d}} \}$ are perfectly correlated. It follows that:
\begin{equation}
    \Delta^2_{\mathrm{rel}} \approx \beta\,\gamma^2 d^2,
    \label{eq:model_variance}
\end{equation}
with
\begin{equation}
    \beta \defn \underset{\mathrm{T}}{\text{Var}}(R_{\text{rel}}(\rho_{\mathrm{T}}; \rho_{\text{opt}}, \lambda(\rho_{\text{opt}}))) .
    \label{eq:beta}
\end{equation}
Note that, if instead we assumed that the states are uncorrelated with each other, we would get $\Delta^2_{\mathrm{rel}} \approx \beta\,\gamma^2 d$.

Eqs. \eqref{eq:model_infidelity} and \eqref{eq:model_variance} define the stochastic model for noise propagation.

\subsection{Estimation of $\alpha$ and $\beta$}
\label{app:estimation_parameters}

In this section we show that $\alpha$ and $\beta$ can be estimated with knowledge of only the distribution of target states and the noise channel properties. 
First of all, we note that both constants depend on the output state $\rho_{\text{opt}}$, which depends on the capability of the circuit to approximate the target state. Since we wish to remove the dependence on the circuit entirely, we use as an approximation $\lambda(\rho_{\text{opt}}) \approx \lambda(\rho_{\mathrm{T}})$. This is justified by the fact that, given a sufficiently expressive circuit, $\rho_{\text{opt}}$ will not be much different from $\rho_{\mathrm{T}}$.
Therefore, using the definitions in Eqs. \eqref{eq:alpha} and \eqref{eq:beta} the parameters can be estimated from the target states as
\begin{align}
    \alpha &\approx \langle R_{\text{rel}}\left(\rho_{\mathrm{T}};\rho_{\mathrm{T}},\lambda(\rho_\mathrm{T})\right) \rangle_\mathrm{T} = \langle R\left(\rho_{\mathrm{T}},\lambda(\rho_\mathrm{T})\right) \rangle_\mathrm{T}, \label{eq:alpha_est}\\
    \beta &\approx \underset{\mathrm{T}}{\text{Var}}(R(\rho_{\mathrm{T}}, \lambda(\rho_\mathrm{T}))) \label{eq:beta_est},
\end{align}
where we can switch the relative infidelity for regular infidelity as we have removed all dependence on the circuit.

Using Eq. \eqref{eq:linear_action} and taking the derivative of $\Lambda(\rho)$ about $\gamma = 0$ we obtain
\begin{equation}
    \frac{d\Lambda(\rho)}{d\gamma}\Bigr|_{\substack{\gamma=0}} = \lambda(\rho) - \rho  \;\;\rightarrow\;\;
    \lambda(\rho) = \frac{d\Lambda(\rho)}{d\gamma}\Bigr|_{\substack{\gamma=0}} + \rho .
\end{equation}
Therefore we can write
\begin{align}
    R(\rho_\mathrm{T}, \lambda(\rho_\mathrm{T})) &= 1 - \text{Tr}\left(\rho_\mathrm{T}\lambda(\rho_\mathrm{T})\right) \nonumber \\
    &= 1 - \text{Tr}\left(\rho_\mathrm{T}\left(\frac{d\Lambda(\rho_\mathrm{T})}{d\gamma}\Bigr|_{\substack{\gamma=0}} + \rho_\mathrm{T} )\right)\right) \nonumber\\
    &= 1 - \text{Tr}\left(\rho_\mathrm{T}\frac{d\Lambda(\rho_\mathrm{T})}{d\gamma}\Bigr|_{\substack{\gamma=0}}\right) - \text{Tr}\left(\rho_\mathrm{T}^2\right) \nonumber \\
    &= - \text{Tr}\left(\rho_\mathrm{T}\frac{d\Lambda(\rho_\mathrm{T})}{d\gamma}\Bigr|_{\substack{\gamma=0}}\right).
\end{align}
Substituting back into Eqs. \eqref{eq:alpha_est} and \eqref{eq:beta_est} we obtain
\begin{align}
    \alpha &\approx -\left\langle\frac{d}{d\gamma}\text{Tr}(\rho_{\mathrm{T}}\Lambda(\rho_{\mathrm{T}}))\Bigr|_{\substack{\gamma=0}}\right\rangle_\mathrm{T} ,\label{eq:alphafinal}\\
    \beta &\approx \underset{\mathrm{T}}{\text{Var}}\left(\frac{d}{d\gamma}\text{Tr}(\rho_{\mathrm{T}}\Lambda(\rho_{\mathrm{T}}))\Bigr|_{\substack{\gamma=0}}\right) \label{eq:betafinal}.
\end{align}

If states can be efficiently sampled from the distribution, the constants $\alpha$ and $\beta$ can be estimated given an exact density matrix simulator. Given a sampled target state, the derivative can be evaluated in practice using finite differences as
\begin{equation}
    \frac{d}{d\gamma}\text{Tr}(\rho_{\mathrm{T}}\Lambda(\rho_{\mathrm{T}}))\Bigr|_{\substack{\gamma=0}} = \underset{\epsilon\rightarrow0}{\text{lim}}\frac{\text{Tr}(\rho_{\mathrm{T}}\Lambda(\rho_{\mathrm{T}}))|_{\substack{\gamma=\epsilon}} - 1}{\epsilon} .
    \label{eq:finitediff}
\end{equation}
The results obtained with Eqs. (\ref{eq:alphafinal}-\ref{eq:finitediff}) for the distribution of 4-qubits real states used in our simulations are shown in Table \ref{tab:1}.
\begin{table}[h]
    \centering
    \begin{ruledtabular}
    \begin{tabular}{ccc}
        Noise channel & $\alpha$ & $\beta$\\
        \midrule
        Phase & 0.888 & 0.00585\\ 
        Amplitude & 1.88 & 0.119\\
        depolarising & 2.78 & 0.0132\\
    \end{tabular}
    \end{ruledtabular}
    \caption{Estimated $\alpha$ and $\beta$ for local noise channels on four qubits, averaged over the real Haar distribution states. 10000 randomly chosen states were used for the numerical estimation.}
    \label{tab:1}
\end{table}

\subsection{Scaling of $\alpha$ with the number of qubits}

For an approximate scaling estimate of $\alpha$ with the number of qubits $N$, we consider simplified case of a target state being a tensor product of $N$ independent identically distributed single-qubit states. With this assumption we can write
\begin{align}
    \alpha &\approx -\frac{d}{d\gamma}\Bigr|_{\substack{\gamma=0}}\text{Tr}[\rho_{\mathrm{T}}\Lambda(\rho_{\mathrm{T}})] \nonumber\\
    &= -\frac{d}{d\gamma}\Bigr|_{\substack{\gamma=0}}\text{Tr}\left[\bigotimes_i \rho_{\mathrm{T}}^{(i)} \Lambda^{(i)}(\rho_{\mathrm{T}}^{(i)})\right] \nonumber\\
    &= -\frac{d}{d\gamma}\Bigr|_{\substack{\gamma=0}}\prod_i \text{Tr}[\rho_{\mathrm{T}}^{(i)} \Lambda^{(i)}(\rho_{\mathrm{T}}^{(i)})] \nonumber\\
    &= -\sum_i \frac{d}{d\gamma}\Bigr|_{\substack{\gamma=0}} \left(\text{Tr}[\rho_{\mathrm{T}}^{(i)} \Lambda^{(i)}(\rho_{\mathrm{T}}^{(i)})]\right) \prod_{j\neq i} \text{Tr}[\rho^{(j)}_{\text{T}} \Lambda^{(j)}|_{\substack{\gamma=0}} (\rho^{(j)}_{\text{T}})] \nonumber \\
    &= -\sum_i \frac{d}{d\gamma}\Bigr|_{\substack{\gamma=0}}\text{Tr}[\rho_{\mathrm{T}}^{(i)} \Lambda^{(i)}(\rho_{\mathrm{T}}^{(i)})] \nonumber\\
    &= \sum_{i=1}^N \alpha^{\text{(i)}} ,
\end{align}
where we defined the single-qubit quantity:
\begin{equation}
    \alpha^{\text{(i)}} := -\frac{d}{d\gamma} \Bigr|_{\substack{\gamma=0}}\text{Tr}[\rho_{\mathrm{T}}^{(i)} \Lambda^{(i)}(\rho_{\mathrm{T}}^{(i)})].
\end{equation}
Since we assumed that the single-qubit product states are identically distributed, these terms are equal for all $i$ and we obtain a scaling of $\alpha \sim O(N)$. Thus, we expect the effect of noise to grow linearly with the number of qubits.
\clearpage

\section{Additional results}
\label{app:plots}

\subsection{VQE for noise levels varying across qubits}

In this subsection we show results for 2-qubit VQE simulations as done in Sec. \ref{sec:two_qubits}, but for the case where the two qubits suffer from very unequal noise levels. We consider the case of one qubit having 10 times the noise of the other one. Analogously to the equal noise results in Fig. \ref{fig:2_qubits_results}, we plot the converged energy, fidelity and concurrence with respect to the noise parameter $\gamma$, which here refers to the most noisy qubit. We do this for the three noise channels (phase, amplitude, depolarising), and focus only on the 4-parameter circuit (Fig. \ref{fig:circuit_2_qubits}(c)). There are two possible configurations: the first is where the most noisy qubit is the top qubit in Fig. \ref{fig:circuit_2_qubits}(c), which we denote as circuit 0; the second is where the bottom qubit is more noisy, which we denote as circuit 1.

\begin{center}
\begin{figure}[!htb]
    \centering
    \includegraphics[width=\textwidth]{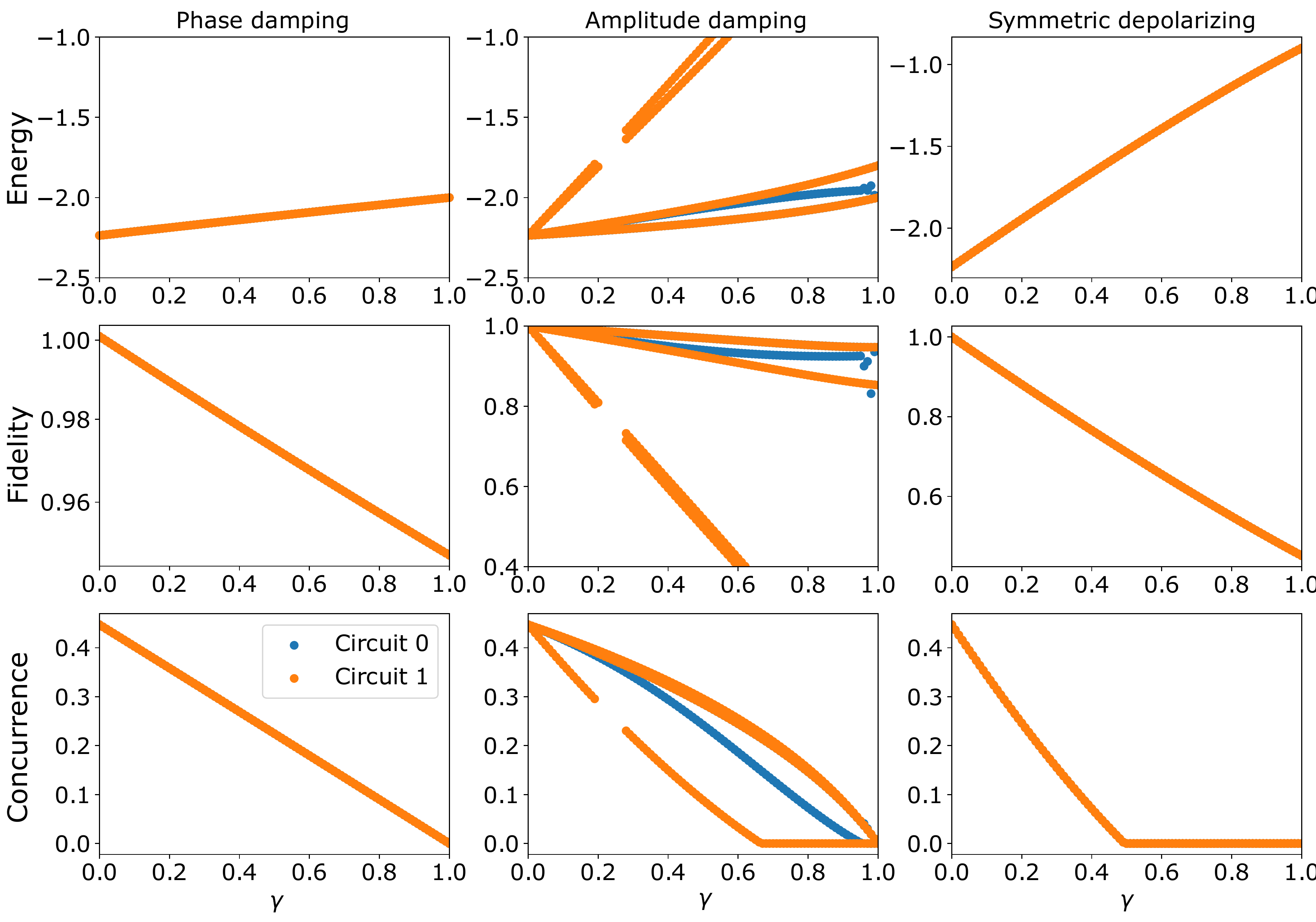}
    \caption{Energy, fidelity and concurrence for the optimal energy minima found for the 2-qubit 4-parameter circuit in Fig. \ref{fig:circuit_2_qubits}c, and for three different types of noise. In these simulations one qubit has a noise level 10 times less than the other: in circuit 0 the first qubit is more noisy, in circuit 1 the second qubit. The columns represent different types of noise channels, the rows the different measures of states quality.}
\label{fig:my_label}
\end{figure}
\end{center}

The results are presented in Fig. \ref{fig:my_label}, and show that for phase and symmetric depolarising noise there is no difference between the two circuits, while for amplitude damping the difference is substantial. In particular, the local minima branch in very distinct directions in the two circuits. This indicates that the degree to which amplitude damping noise breaks the parameter degeneracies, leading to different local minima with increasing $\gamma$, is sensitive to the relative strength of the noise on the different qubits.

\onecolumngrid
\clearpage
  
\twocolumngrid

\subsection{Relative infidelity for different noise channels}

We show the results for relative infidelity for amplitude damping and symmetric depolarising noise, for the target state optimization experiments. The results are similar to those for phase damping noise, presented in the main text (Fig. \ref{fig:different_minima}). There is a good agreement between the stochastic model and the numerical results for low noise parameter values. The model follows more closely the non-reoptimised results, which in the symmetric depolarising case match almost exactly the reoptimised results. Furthermore, the agreement improves with the number of layers, indicating that the assumptions of random intermediate states are more appropriate for deeper circuits.

\onecolumngrid

\begin{center}
\begin{figure}[!htb]
\centering
\begin{subfigure}{.5\textwidth}
  \centering
  \includegraphics[width=\textwidth]{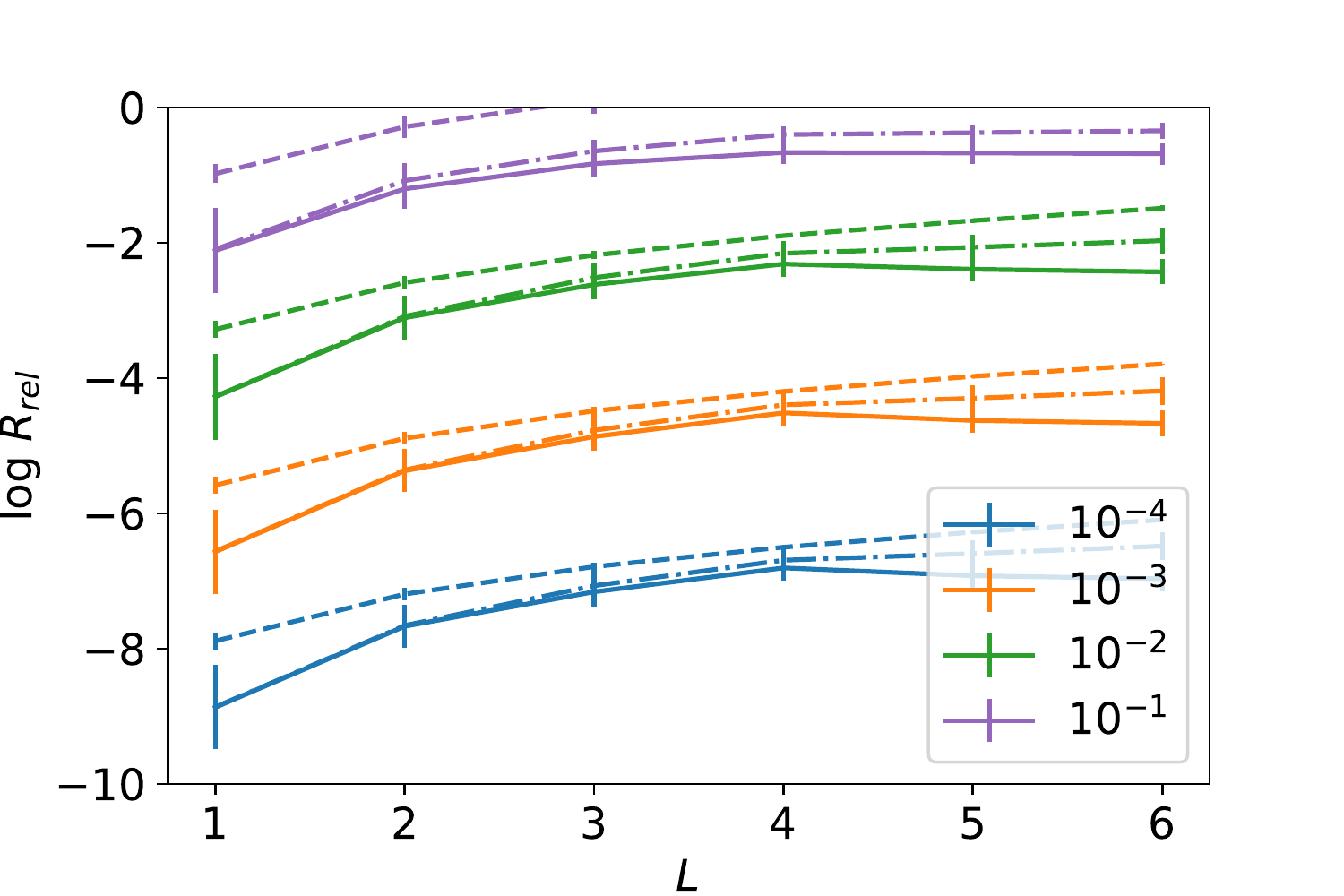}
  \caption{Amplitude damping noise}
\end{subfigure}%
\begin{subfigure}{.5\textwidth}
  \centering
  \includegraphics[width=\textwidth]{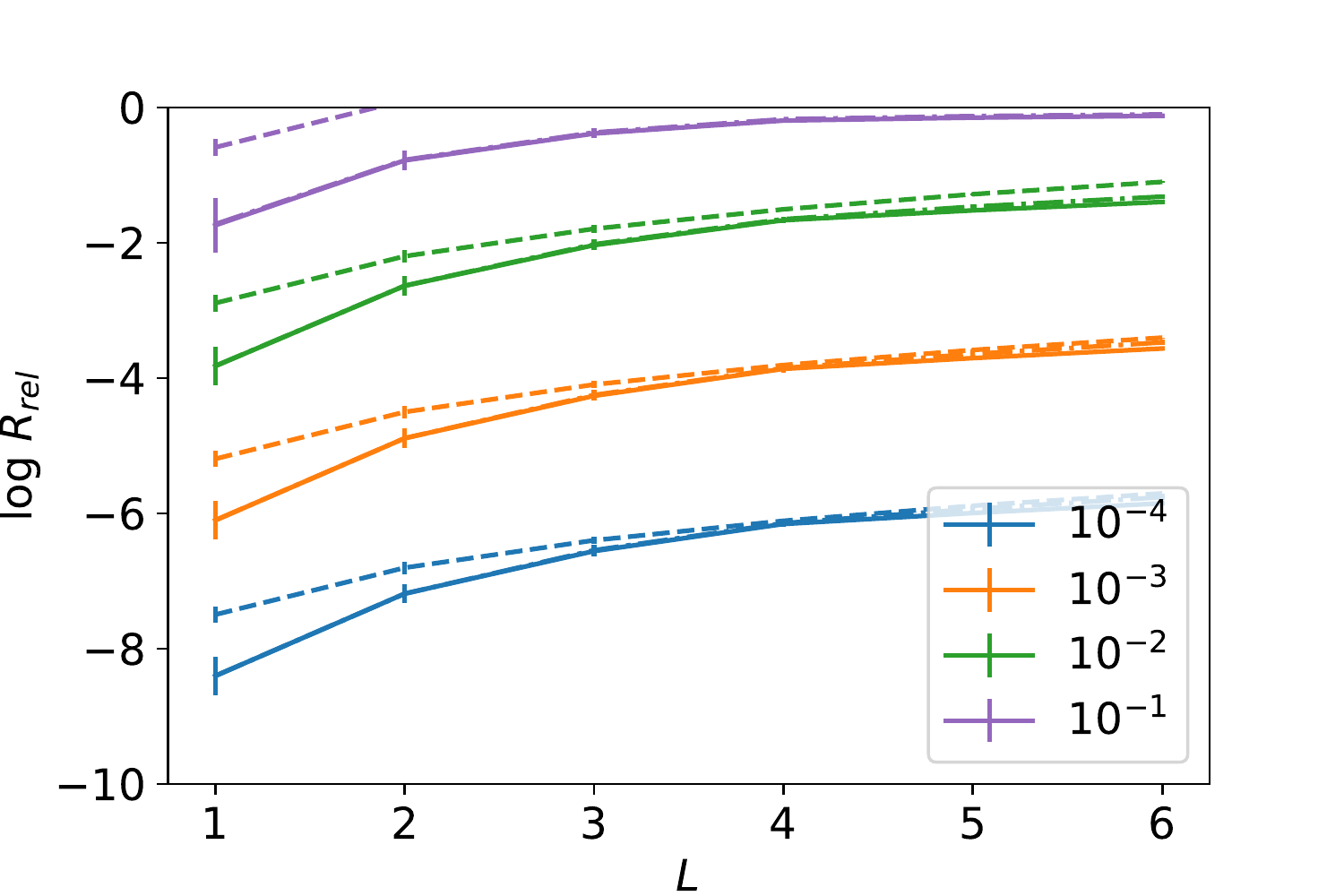}
  \caption{Symmetric depolarising noise}
\end{subfigure}
\caption{Relative infidelity vs layers, for a) amplitude damping noise and b) symmetric depolarising noise, for different values of the noise level $\gamma$. The solid curves are for noise-aware reoptimised parameters, the dash-dotted curves are for parameters optimised in the absence of noise (non-reoptimised), and the dashed curves indicate the results for the linear noise model. Each point shows the average over 1000 target states, and the verical bars at each point indicate the standard deviation.}
\label{fig:test}
\end{figure}
\end{center}

\clearpage

\twocolumngrid

\subsection{Noise-induced state transitions for further random states}

\begin{center}
\begin{figure}[!htbp]
\centering
    \includegraphics[width=0.33\textwidth]{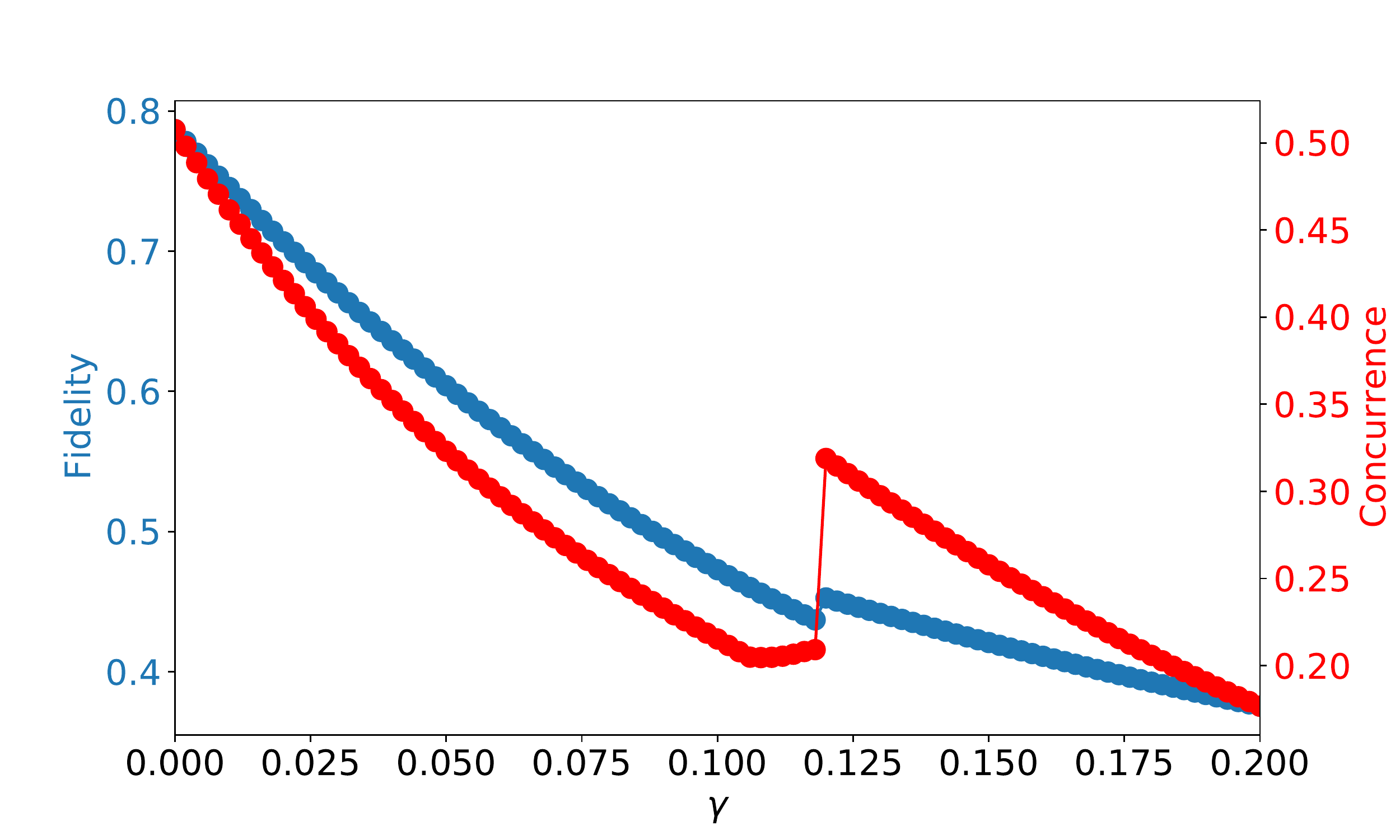}%
    \includegraphics[width=0.33\textwidth]{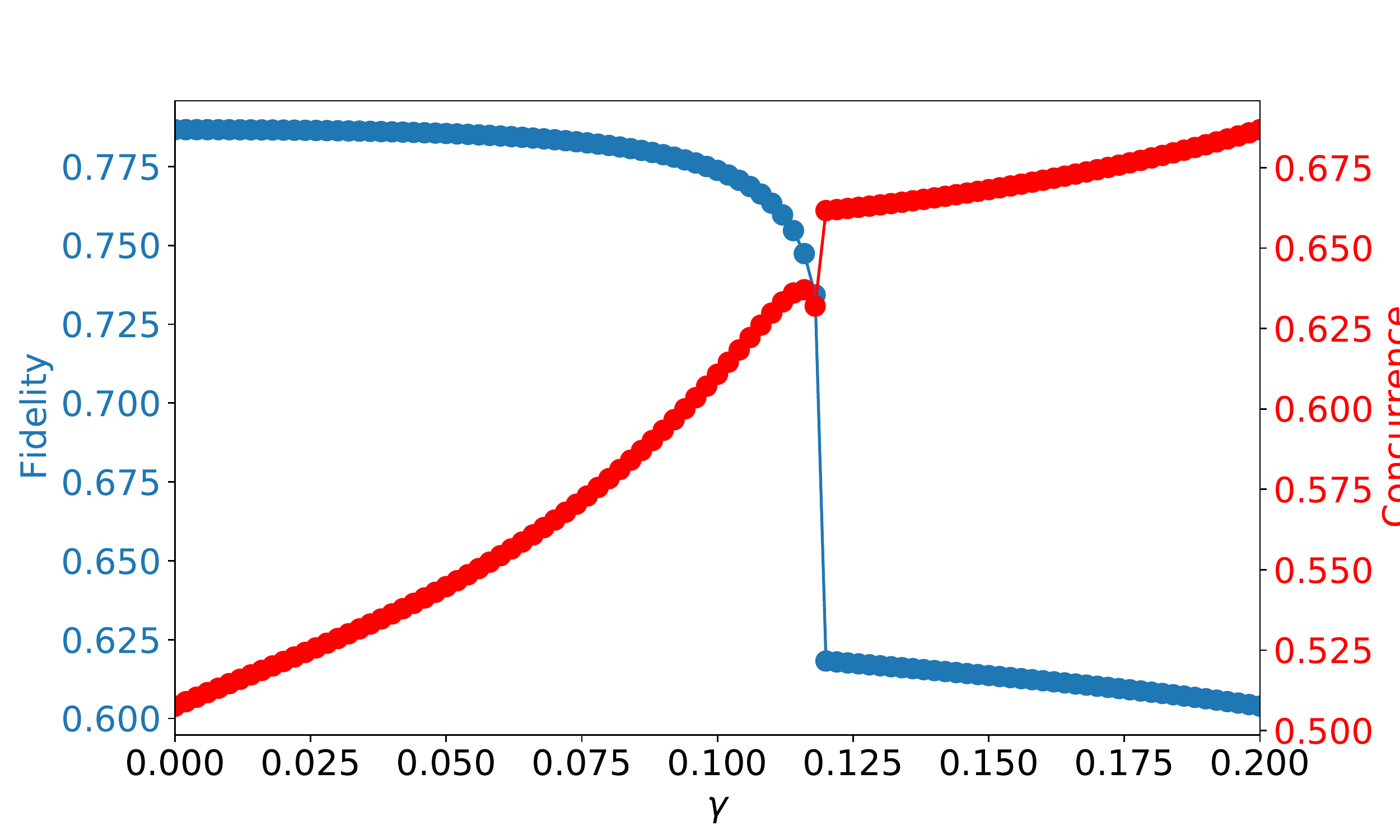}%
    \includegraphics[width=0.4\textwidth]{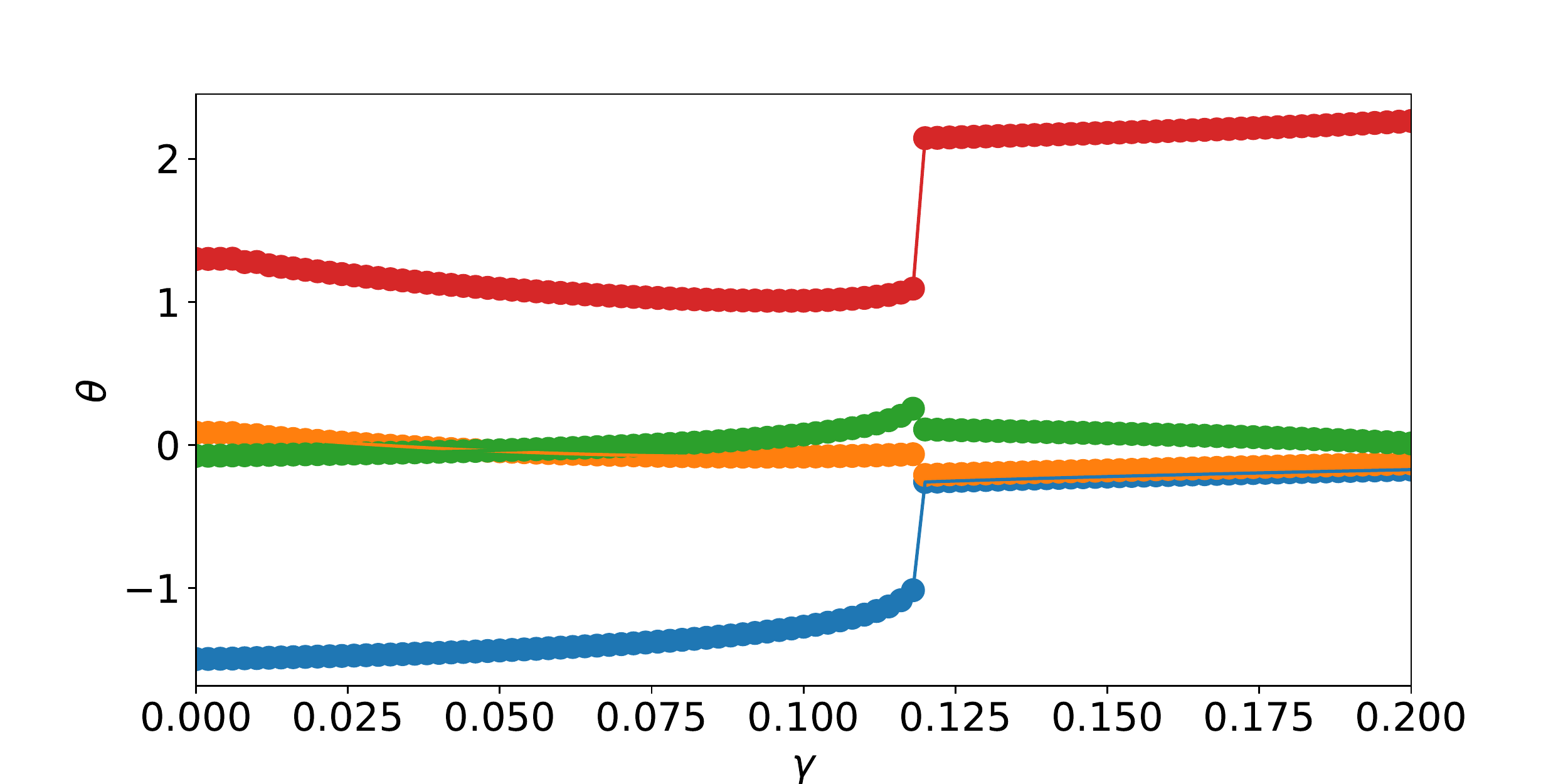}

    \includegraphics[width=0.33\textwidth]{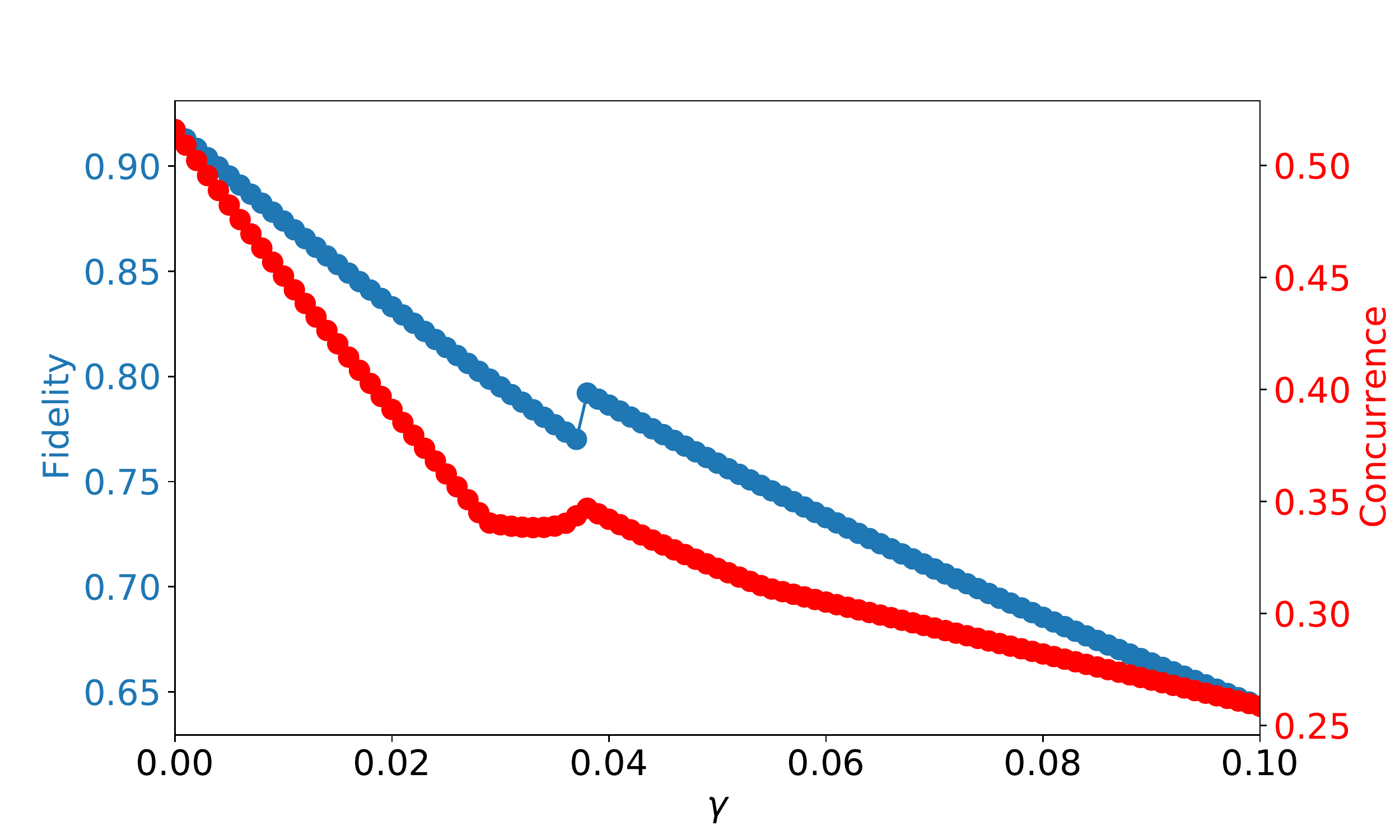}%
    \includegraphics[width=0.33\textwidth]{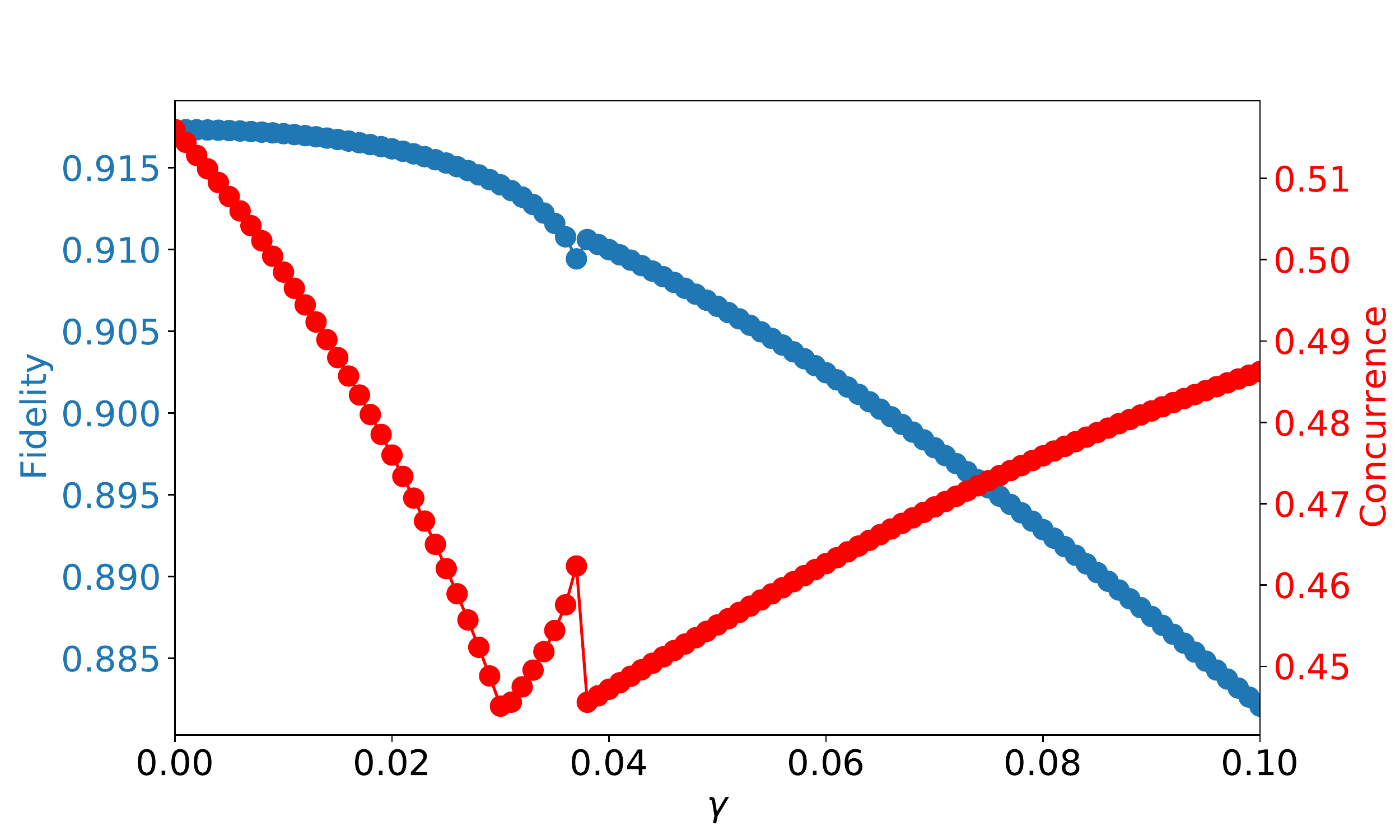}%
    \includegraphics[width=0.4\textwidth]{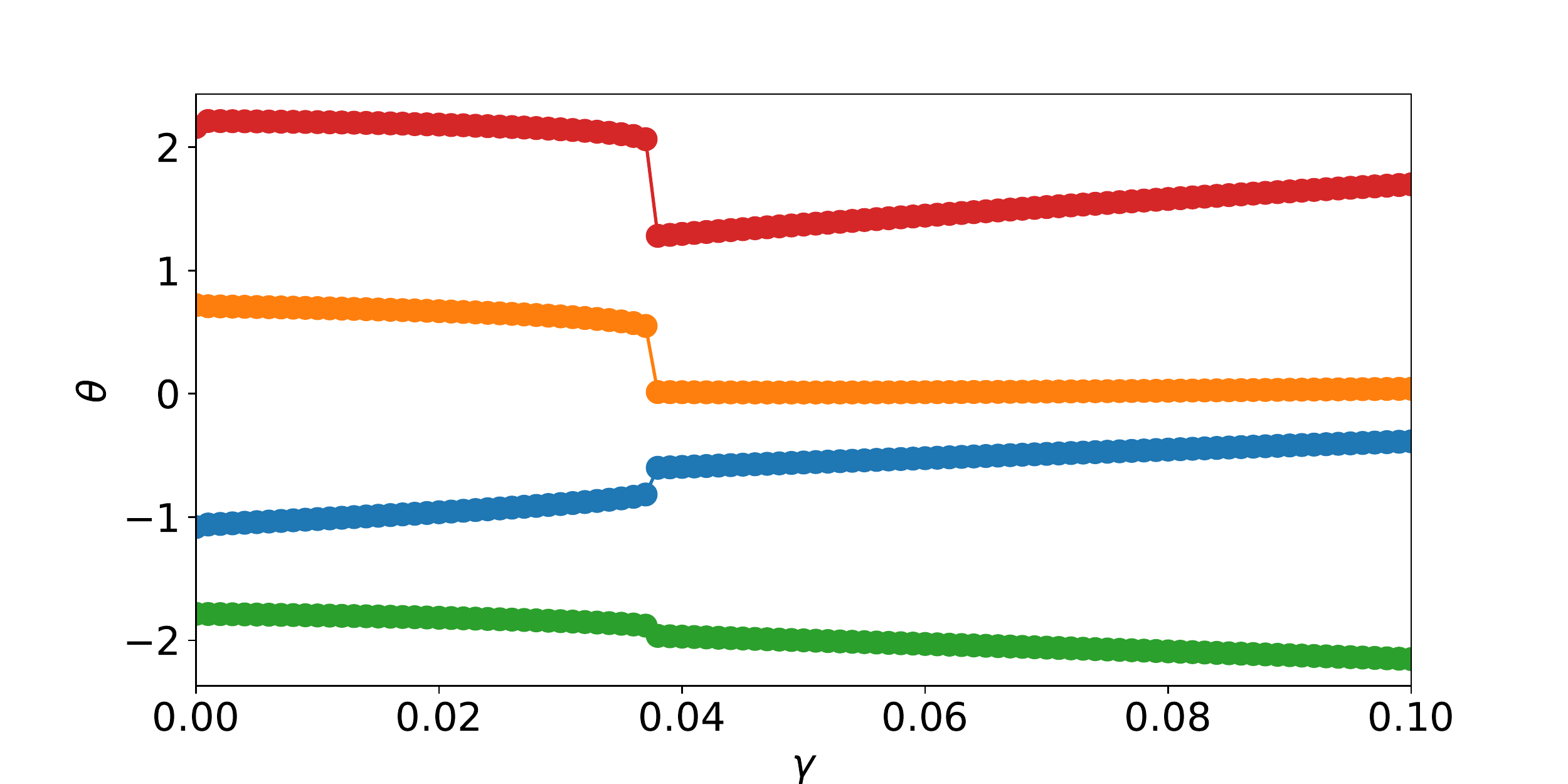}
    
    \includegraphics[width=0.33\textwidth]{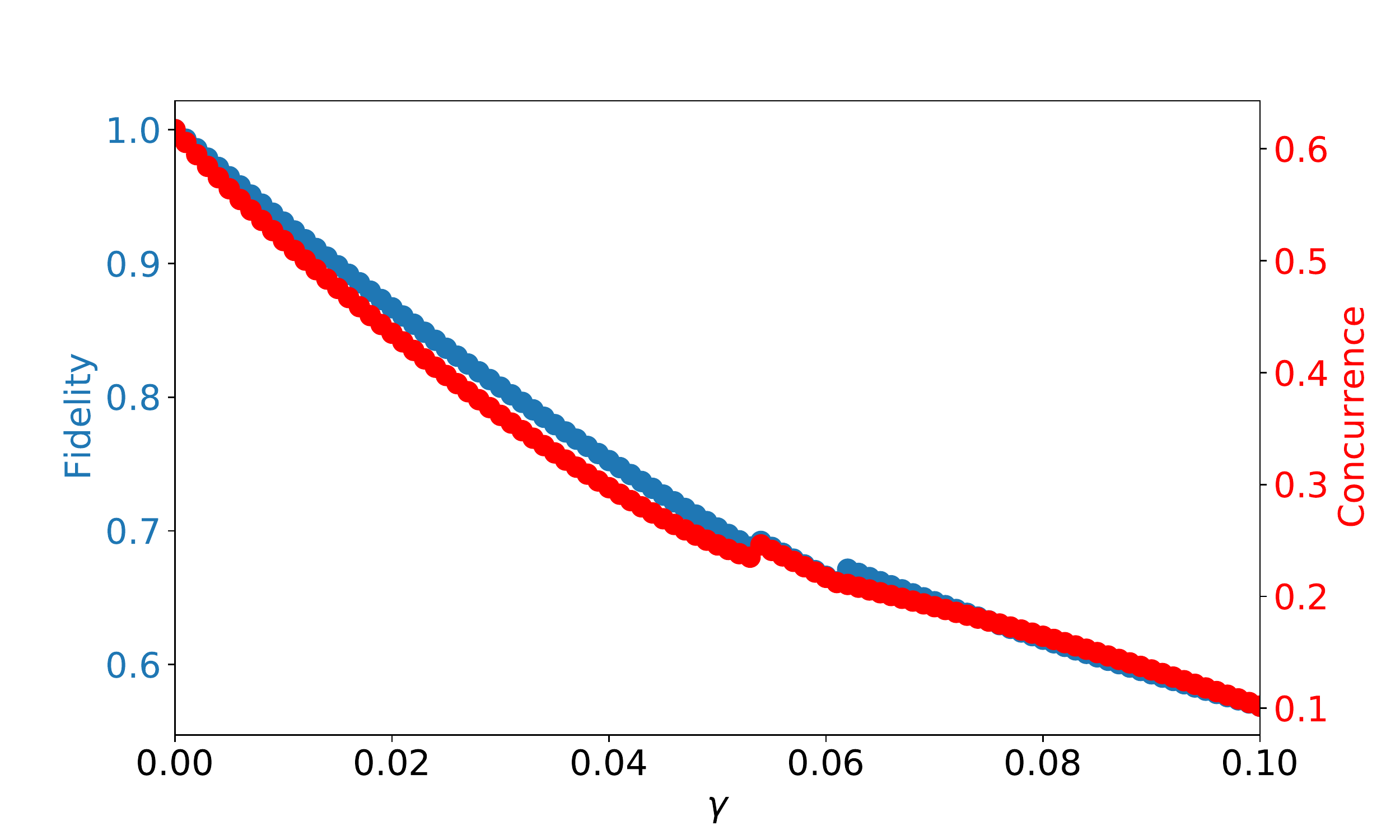}%
    \includegraphics[width=0.33\textwidth]{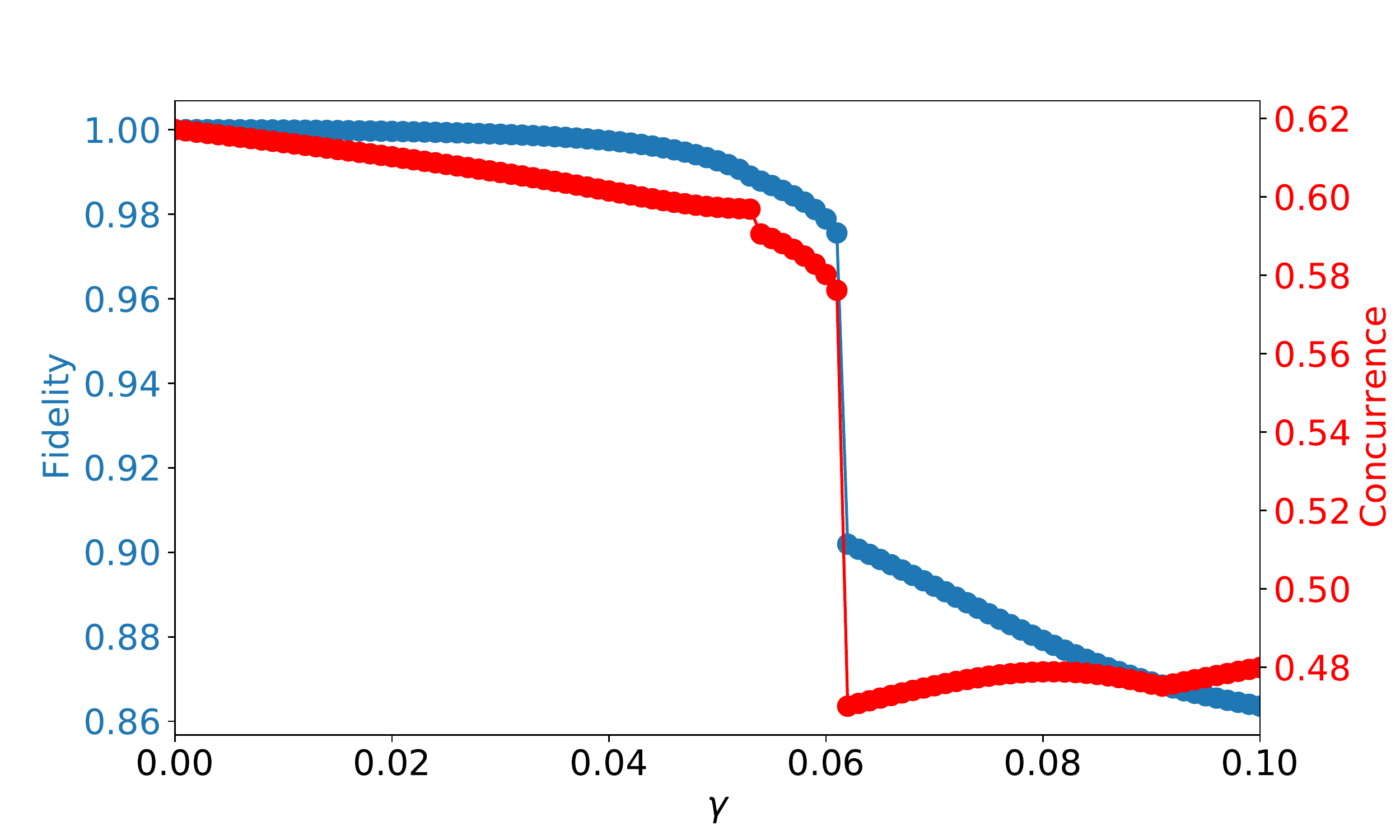}%
    \includegraphics[width=0.4\textwidth]{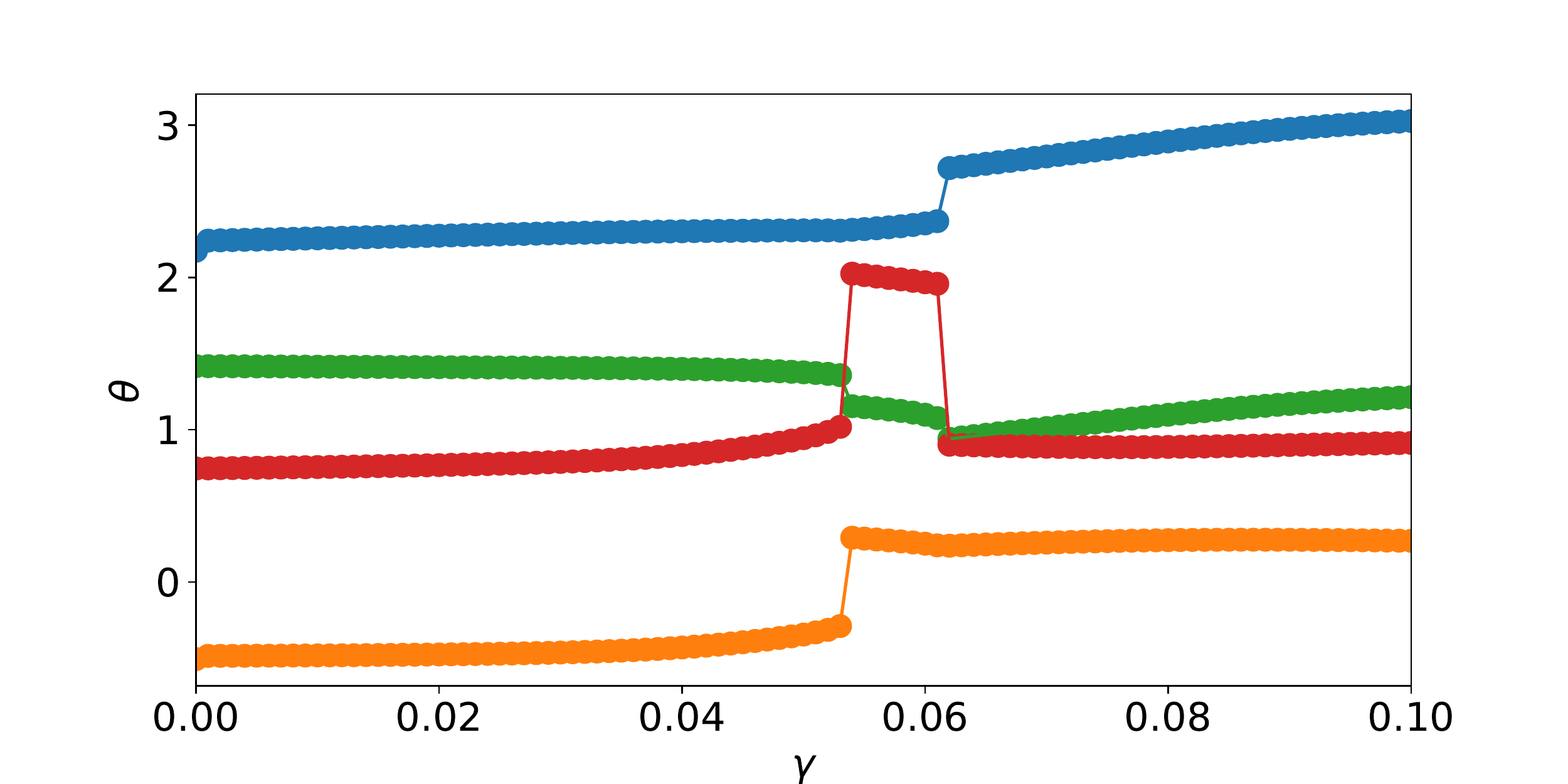}
    
    \includegraphics[width=0.33\textwidth]{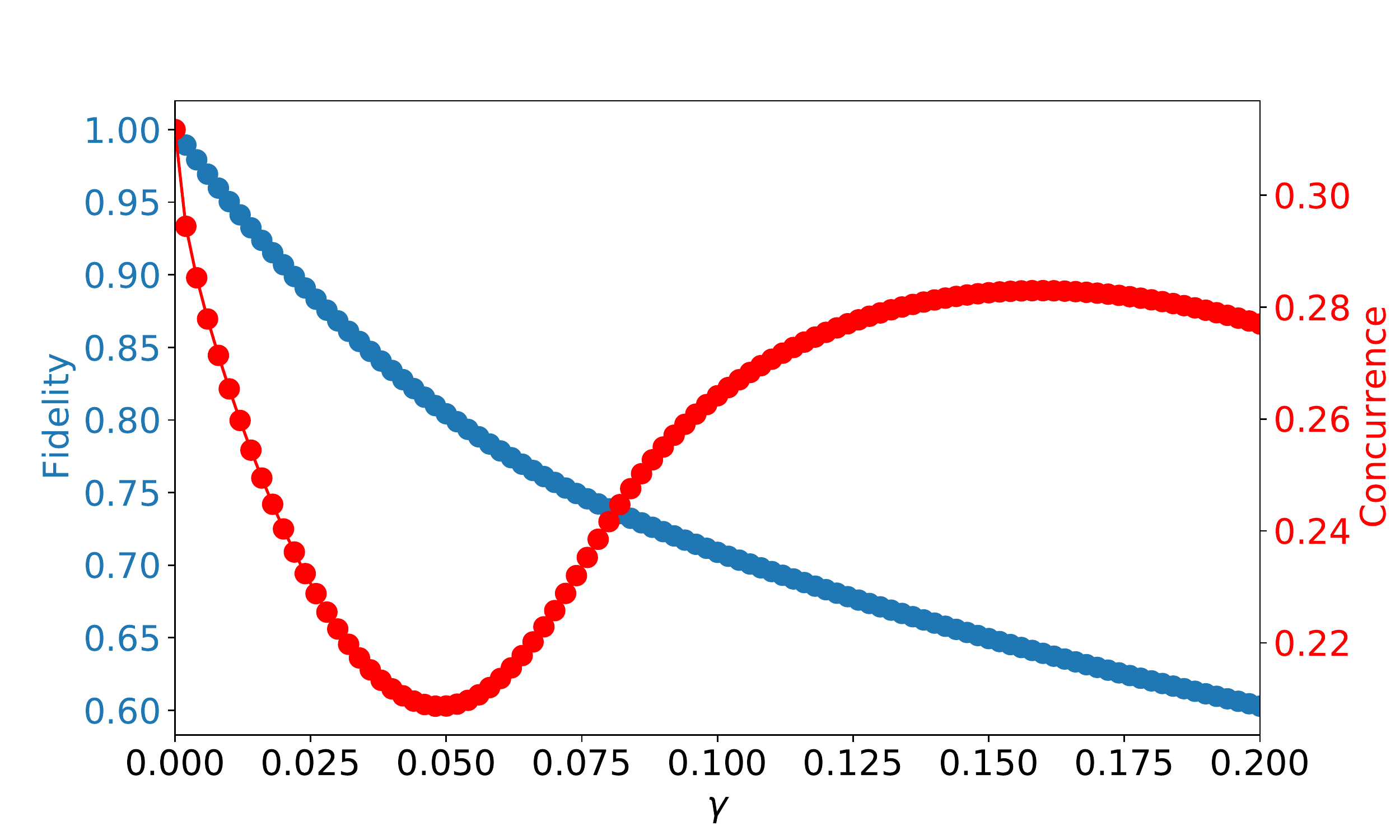}%
    \includegraphics[width=0.33\textwidth]{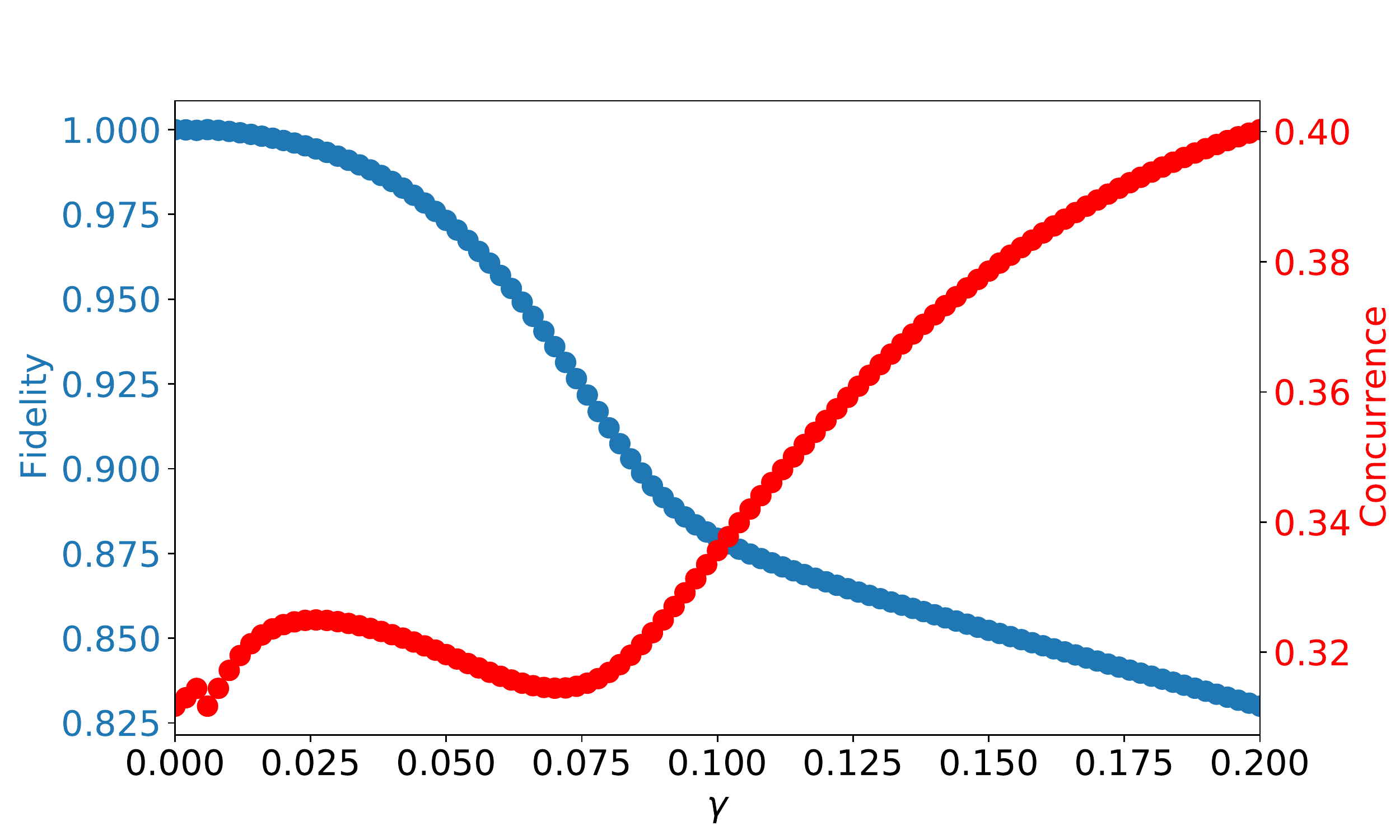}%
    \includegraphics[width=0.4\textwidth]{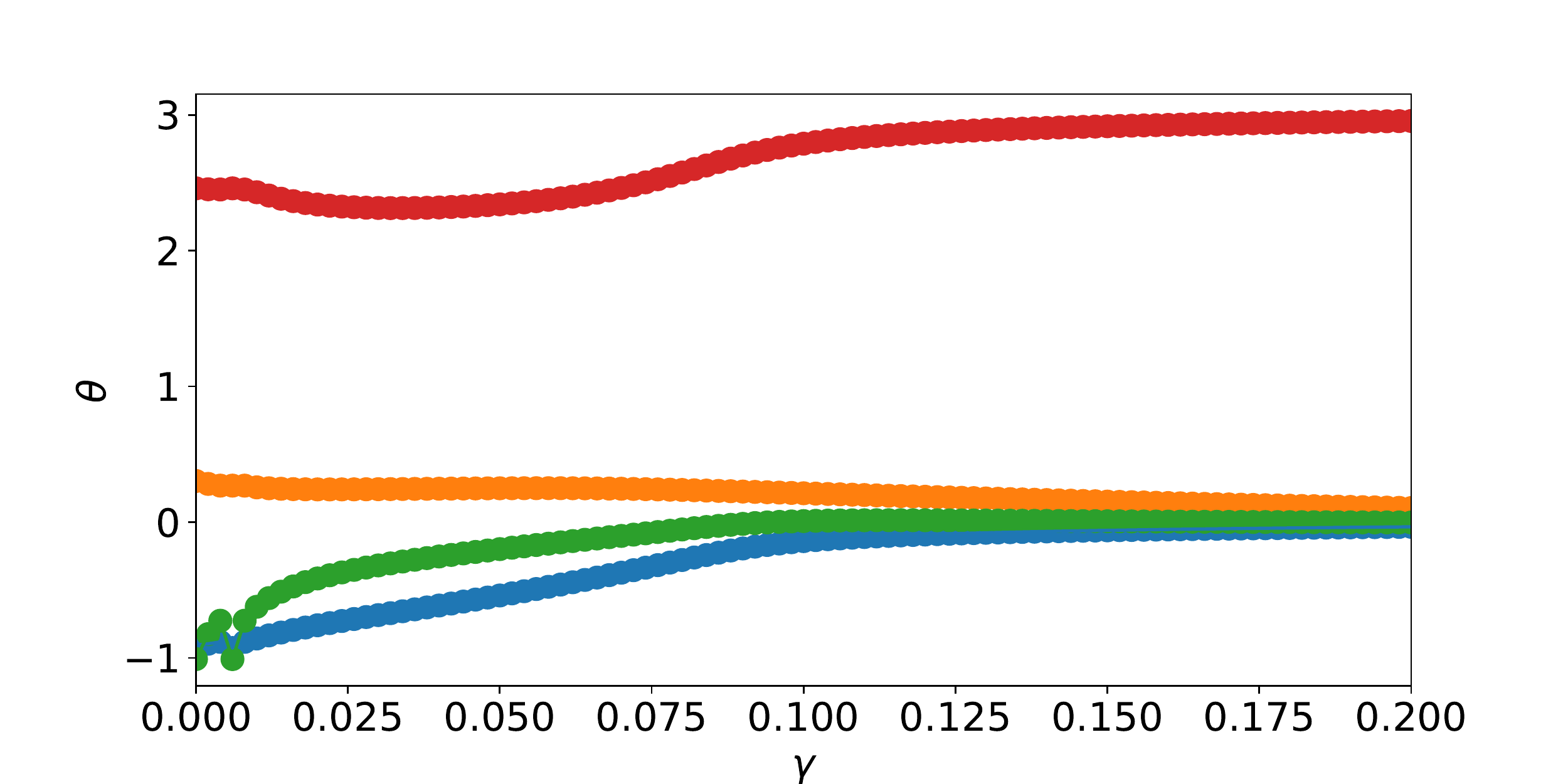}
    \caption{Evidence of noise-induced state transitions for different random states. Various measures are plotted at convergence for a range of values of the phase noise parameter $\gamma$. The plotted data are described in more detail in fig. \ref{fig:different_minima}. Left panels: fidelity and concurrence at convergence, evaluated with noise; centre panels: fidelity and concurrence evaluated without noise, but for the converged optimized angles at every noise level; right panels: value of four of the circuit rotation parameters at convergence. The first two rows have a circuit of depth $L=3$, the last two have $L=4$.}
    \label{fig:more_different_minima}
\end{figure}
\end{center}

In Fig. \ref{fig:different_minima} we show how fidelity and concurrence as function of $\gamma$ abruptly change slope at at threshold $\gamma$ value for one randomly chosen target state. To show that this behavior is general, in Fig. \ref{fig:more_different_minima} we present analogous results for four more randomly chosen target states.
We also consider different circuit depths. The transitions are overall very different to one another, at times discontinuous in their fidelity and at other times smooth and barely detectable. In these latter cases, the concurrence acts as a clearer indicator for such transitions. The varied phenotype of transitions suggests that the phenomenon is complex and depends heavily on the circuit and the chosen target state. Indeed, in some situations (see the last row of Fig. \ref{fig:more_different_minima}) no sharp transitions are observed, and both measures of state quality vary smoothly. The results illustrate that generally one always observes critical noise level thresholds, and that these can be either abrupt and discontinuous or else smoothed out in a continuous way. 

\onecolumngrid

\twocolumngrid

\bibliography{Noisy_VQE}

\end{document}